%% file: conference_101719.tex
\documentclass[conference]{IEEEtran}
\IEEEoverridecommandlockouts
\usepackage{cite}
\usepackage{amssymb,amsfonts}
\usepackage{algorithmic}
\usepackage{textcomp}
\usepackage{tcolorbox}
\usepackage{colorprofiles}
\usepackage{amsmath}
\usepackage{amsthm}
\usepackage{dsfont}
\usepackage{listings}
\usepackage{svg}
\usepackage{float}
\usepackage{xcolor}
\usepackage{pgfplots}
\pgfplotsset{compat=1.18}
\usepackage[braket, qm]{qcircuit}
\usepackage{graphicx}
\usepackage{subcaption}
\usepackage{adjustbox}
\usepackage{tikz}
\input{tikzit.sty}
\input{sample.tikzstyles}
\input{sample.tikzdefs}
\usetikzlibrary{positioning, arrows, calc, math, angles, quotes}
\usepackage{blochsphere}
\usepackage{placeins}
\usepackage{listings}
\usepackage{xcolor}
\usepackage{hyperref}
\usepackage{balance}

\colorlet{punct}{red!60!black}
\definecolor{background}{HTML}{EEEEEE}
\definecolor{delim}{RGB}{20,105,176}
\colorlet{numb}{magenta!60!black}

\lstdefinelanguage{json}{
    basicstyle=\normalfont\ttfamily,
    stepnumber=1,
    numbersep=8pt,
    showstringspaces=false,
    breaklines=true,
    frame=lines,
    backgroundcolor=\color{background},
    literate=
     *{0}{{{\color{numb}0}}}{1}
      {1}{{{\color{numb}1}}}{1}
      {2}{{{\color{numb}2}}}{1}
      {3}{{{\color{numb}3}}}{1}
      {4}{{{\color{numb}4}}}{1}
      {5}{{{\color{numb}5}}}{1}
      {6}{{{\color{numb}6}}}{1}
      {7}{{{\color{numb}7}}}{1}
      {8}{{{\color{numb}8}}}{1}
      {9}{{{\color{numb}9}}}{1}
      {:}{{{\color{punct}{:}}}}{1}
      {,}{{{\color{punct}{,}}}}{1}
      {\{}{{{\color{delim}{\{}}}}{1}
      {\}}{{{\color{delim}{\}}}}}{1}
      {[}{{{\color{delim}{[}}}}{1}
      {]}{{{\color{delim}{]}}}}{1},
}

\usetikzlibrary{positioning}

\newcommand{\rulesep}{\unskip\ \vrule\ }

\theoremstyle{definition}
\newtheorem{definition}{Definition}[section]
\theoremstyle{remark}
\newtheorem{example}{Example}

\begin{document}

\title{SQL2Circuits: Estimating Cardinalities, Execution Times, and Costs for SQL Queries with Quantum Natural Language Processing}

\input{authors}

\maketitle

\begin{abstract}
Recent advances in quantum computing have led to progress in exploring quantum applications across diverse fields, including databases and data management. This work presents a quantum machine learning model that tackles the challenge of estimating metrics, such as cardinalities, execution times, and costs, for SQL queries in relational databases. Precise estimations are crucial for the query optimizer to optimize query processing in relational databases efficiently. Our proposed quantum machine learning model consists of a novel query encoding mechanism, which maps SQL queries into high-dimensional Hilbert spaces using grammatical representations of the queries. The encoding mechanism translates SQL queries into parameterized quantum circuits, forming the core of the quantum machine learning model. The parameters in this model are tuned using standard quantum machine learning techniques. This encoding was first developed in quantum natural language processing (QNLP), and this work demonstrates its natural application in database optimization. Because the encoding mechanism is mathematically robust, the quantum machine learning model is also explainable, allowing us to draw a one-to-one correspondence between the elements in SQL queries and the model's parameters. The method is also scalable because it consists of multiple circuits, and we train and evaluate the model with hundreds of queries. Compared to previous research, our model achieves high accuracy, supporting the results obtained in the original QNLP research. We extend the previous QNLP work by adding 4-class and 8-class classification tasks and comparing the cardinality estimation results with those from state-of-the-art databases. This model shows a competitive accuracy compared to the estimators in the databases. We theoretically analyze the quantum machine learning model by calculating its expressibility and entangling capabilities, which have been suggested as representative metrics describing the model's trainability and learnability features.
\end{abstract}


\input{sections/introduction.tex}
\input{sections/prerequisites.tex}
\input{sections/encoding.tex}
\input{sections/training_main.tex}
\input{sections/evaluation.tex}
\input{sections/analysis_of_model.tex}
\input{sections/conclusion_future_work.tex}

\FloatBarrier
\balance
\bibliographystyle{IEEEtran}
\bibliography{ref.bib}

\end{document}

%% file: sample.tikzstyles
\tikzstyle{box}=[fill=white, draw=black, shape=rectangle]

\tikzstyle{dashed edge}=[-, thick, dashed]

%% file: authors.tex
\author{\IEEEauthorblockN{Valter Uotila}
\IEEEauthorblockA{University of Helsinki \\
first.last@helsinki.fi}
}

%% file: sections/introduction.tex
\section{Introduction}

Quantum computing has made steady progress in recent years, and there is a growing interest in its practical applications. Improvements in quantum hardware, enhanced algorithms, and advanced error correction techniques have significantly increased the potential of quantum computing to tackle specific problems, such as optimization and specialized tasks in machine learning. While significant challenges remain, these developments have brought quantum computing closer to real-world use cases in research and industry.

Quantum computing applied to databases is an emerging research domain \cite{10.1145/3555041.3589404, Groppe22QDM, Schonberger_2022, Uotila_2022}. Quantum computing has demonstrated its potential utility in various database problems, which include multiple query optimization \cite{Trummer_Koch_2015}, transaction scheduling \cite{Bittner_Groppe_2020, OJCC_2020v7i1n01_Bittner, Groppe22QDM}, virtual machine allocation \cite{Uotila_Lu_2023}, index tuning \cite{Gruenwald_Winker_Caliyilmaz_Groppe_Groppe_2023}, and especially join order optimization \cite{schonberger2023ready, Schonberger_Trummer_Mauerer, 10.1145/3579142.3594299,uotila2025leftdeepjoinorderselection}. The primary methods have been unconstrained binary optimization and quantum machine learning.

In this work, our objective is to apply a novel quantum machine learning model to estimate metrics of SQL queries. Given an SQL query, the method estimates its cost, execution time, or cardinalities without executing the query. For example, given a SQL query, an execution time estimation could be 3 seconds, cardinality estimation would be 1000 rows, and cost would be 1729 database-specific cost units. Our method generalizes to any other metrics we can assign to queries, but we show the results for these three cases. The estimations are useful for countless applications in query optimization, such as join order selection. The problem has been tackled from different perspectives. The existing work includes synopsis-based \cite{Wu_Chi_Zhu_Tatemura_Hacigumus_Naughton_2013}, sampling-based  \cite{conf/sigmod/QiuWYLWZ21}, and machine learning-based methods \cite{Akdere_Cetintemel_Riondato_Upfal_Zdonik_2012, Ganapathi_Kuno_Dayal_Wiener_Fox_Jordan_Patterson_2009, Tozer_Brecht_Aboulnaga_2010, Xiong_Chi_Zhu_Tatemura_Pu_Hacigumus_2011}.

To address the problem of estimating metrics with quantum computing, we convert the estimation problem into a query classification task. Given a predefined set of classes, the method decides which class a given query belongs to without executing it. Each class is associated with a range of values, and the ranges provide query estimations. For example, execution time ranges could be 0 seconds to 5 seconds (class 1) and from 5 seconds to 10 seconds (class 2). Similarly, cardinalities could be 0-1000 rows (class 1) and 1000-2000 rows (class 2). 

The complete quantum machine learning model is built and used as follows. We have created a synthetic dataset of queries that query data from the IMDB database \cite{IMDb}. The generated queries are executed on a PostgreSQL relational database, and the training and test data containing execution times, cardinalities, and costs are collected. The next step is to apply the novel encoding mechanism, which we describe in detail in the coming sections. The encoding mechanism translates the queries into parametrized quantum circuits. The parameters are optimized so that the circuits, when executed on a quantum computer, classify the SQL queries into two, four, or eight classes corresponding to the estimations. 

Specifically, we highlight the new advantages that the proposed quantum machine learning model offers for this database task compared to many classical and quantum algorithms. These aspects motivate the usage and research of this model for the database-related problem:
\begin{itemize}
    \item Our quantum machine learning model is unified: we can use the same model (with the same queries, structure, circuits, layers, and hyperparameters) with different parameters to predict execution times, cardinalities, costs, and any other relevant metrics for which the model has been trained.
    \item The method is database-agnostic and can be applied to relational, graph, and document databases, as well as other databases that use grammar-based query languages.
    \item Our model does not depend on queries: it can handle any query that a database can execute. Many previous works have assumed that the estimated queries do not contain certain advanced filtering conditions.
    \item Because our model is based on category theoretical transformations, as the other QNLP-based models \cite{Lorenz_Pearson_Meichanetzidis_Kartsaklis_Coecke_2021, Miranda_Yeung_Pearson_Meichanetzidis_Coecke_2021}, it becomes explainable: we have a one-to-one correspondence between the query elements (tables, attributes, joins, keywords, filtering conditions) and the parameters. Moreover, it admits many other aspects of explainable AI that have been proved in previous research \cite{tull2024compositionalinterpretabilityxai}.
    \item Our model can learn from minimal data and generalize well: with just 400 training queries, we achieve high accuracy in predicting 300 validation and test queries.
    \item A new iterative training pipeline: The pipeline addresses that quantum machine learning models are notoriously challenging to train \cite{bowles2024betterclassicalsubtleart}. It also demonstrates how the model is adaptive to the updates in the SQL queries and training data.
    \item Natural language does not necessarily follow any grammar, while database queries always do. Thus, we argue that the model is better suited for the selected database problem than for natural language processing.
\end{itemize}

The main contributions of this paper are summarized as follows:
\begin{enumerate}
    \item We develop a mathematically robust and explainable quantum machine learning model that encodes SQL queries as quantum circuits and estimates various metrics for the queries. Data encoding is developed formally based on category theory.
    \item We present a comprehensive quantum machine learning pipeline with a novel iterative training method and implement the previously mentioned enhancements over earlier versions of this quantum machine learning model.
    \item The cardinality estimations are compared to the results from state-of-the-art open-source and commercial databases. The results indicate that the model can perform well, considering that quantum machine learning is usually developed at a scale where this comparison is not even feasible.
    \item We provide a theoretical analysis of the expressibility and entangling capabilities of the model's circuits. These metrics have not been previously considered for this type of quantum machine learning model.
\end{enumerate}

The paper's outline is as follows. First, we briefly review the basics of the mathematical background that is used to develop the model. We then apply the mathematically robust process of encoding SQL queries as parametrized circuits and optimizing these circuit parameters. Then, we improve the training pipeline and present the experimental results. Finally, we show the results from theoretical analysis. The framework and datasets can be accessed online \cite{repo}.

%% file: sections/prerequisites.tex
\section{Background}
This work resides at the intersection of quantum computing, database research, and category theory. Category theory provides a mathematically precise formal model to describe the novel encoding mechanism, making the quantum machine learning model more explainable \cite{tull2024compositionalinterpretabilityxai}. We assume that the reader is familiar with the basics of quantum computing \cite{Nielsen_Chuang_2010}. A recent introduction to quantum computing from the database optimization perspective is presented in \cite{10.1145/3555041.3589404}. The comprehensive introduction to practical category theory is \cite{Coecke_Kissinger_2017}. Most of the quantum natural language processing research \cite{Lorenz_Pearson_Meichanetzidis_Kartsaklis_Coecke_2021, Miranda_Yeung_Pearson_Meichanetzidis_Coecke_2021, kartsaklis2021lambeq} also introduces the basics of category theory. The abstract category theory is presented in \cite{Riehl_2017, MacLane_1978}.


\input{sections/discocat.tex}

%% file: sections/discocat.tex

The theoretical background for the quantum machine learning models in QNLP and this article is based on category theory \cite{Riehl_2017, Spivak_2014}, which can be viewed as an alternative to set theory. Category theory was initially developed to explain mathematics, making it a conceptually powerful tool. It also provides a precise and visual approach to express countless abstract and practical concepts, such as SQL queries, their grammatical representations, and the corresponding circuits. As far as we know, this is the first application of category theory to model SQL queries. 

This paper's applied category theoretical approach is Distributional, Compositional, and Categorical (DisCoCat) modeling. DisCoCat modeling has been used primarily in quantum natural language processing \cite{Lorenz_Pearson_Meichanetzidis_Kartsaklis_Coecke_2021, Miranda_Yeung_Pearson_Meichanetzidis_Coecke_2021,duneau2024scalable}. Also, quantum mechanics and quantum computing are formalized with this type of theory \cite{Coecke_Kissinger_2017}. To make this paper self-contained, we briefly recall the necessary definitions from category theory \cite{MacLane_1978, Riehl_2017}. While we are developing the mapping from the SQL queries to the quantum machine learning model, we will constantly refer back to these concepts. 

\begin{definition}{\textbf{Category}.}\label{def:category}
A category $\mathcal{C}$ consists of a collection of (abstract) objects, denoted by $\mathrm{Obj}(\mathcal{C})$, and a collection of morphisms, denoted by $\mathrm{Hom}(\mathcal{C})$. The objects and morphisms satisfy the following properties.
\begin{itemize}
    \item Every morphism $f$ has specified domain and codomain objects.
    \item Every object $X \in \mathrm{Obj}(\mathcal{C})$ has an associated identity morphism $\mathrm{id}_{X} \colon X \to X$.
    \item The category $\mathcal{C}$ has a special composition operation for morphisms. Moreover, the composition operation respects identity morphisms and is associative.
\end{itemize}
\end{definition}

This work uses three mappings to construct the encoding process from SQL queries to our quantum machine learning model. Formally, these mappings are defined as functors, which are mappings between categories.

\begin{definition}{\textbf{Functor}.}\label{def:functor}
Let $\mathcal{C}$ and $\mathcal{D}$ be categories. A functor $F \colon \mathcal{C} \to \mathcal{D}$ consists of a mapping between the objects and a mapping between morphisms of the categories $\mathcal{C}$ and $\mathcal{D}$.
\begin{itemize}
    \item For every object $X \in \mathrm{Obj}(\mathcal{C})$, there is an object $F(X) \in \mathrm{Obj}(\mathcal{D})$.
    \item For every morphism $f \colon X \to Y \in \mathcal{C}$, there exists morphisms $F(f) \colon F(X) \to F(Y) \in \mathcal{C}$ so that the domain $F(X)$ (and codomain $F(Y)$) of $F(f)$ are equal to the domain $X$ (and codomain $Y$) of $f$ when $F$ is applied to $X$ (and $Y$).
\end{itemize}
Furthermore, we assume that the following functoriality axioms are satisfied.
\begin{itemize}
    \item $F(g \circ f) = F(g) \circ F(f)$ for any composable pair $f$ and $g$ in the category $\mathcal{C}$. 
    \item $F(\mathrm{id}_{X}) = \mathrm{id}_{F(X)}$ for $X \in \mathrm{Obj}(\mathcal{C})$.
\end{itemize}
\end{definition}

Finally, in practical applications of category theory, we often need to assume that categories have additional structure. One of these additional structures is an abstract tensor product. A category with a tensor product is called a monoidal category. 

\begin{definition}{\textbf{Monoidal category}.}\label{def:monoidal_category}
A monoidal category $\mathcal{C}$ is a category equipped with a (bi)functor $\otimes \colon \mathcal{C} \times \mathcal{C} \to \mathcal{C}$ (tensor product) which satisfies the following assumptions. First, there exists an identity object $I \in \mathrm{Obj}(\mathcal{C})$. Second, the tensor product $\otimes$ is associative, meaning that $A \otimes (B \otimes C) \cong (A \otimes B) \otimes C$. Finally, the identity object $I$ satisfies that $I \otimes A \cong A \otimes I \cong A$.
\end{definition}



%% file: sections/encoding.tex
\section{Map SQL queries into parametrized circuits}\label{sec:sqltocirc}
Programming languages and database query languages are well-defined constructions that follow a special set of rules that are defined in their grammars. This section describes the first phase in constructing our quantum machine learning model, the data encoding mechanism, which encodes SQL queries into parametrized quantum circuits using context-free grammar for SQL queries. Intuitively, the encoding first maps SQL queries into the corresponding context-free grammar representation, then to the corresponding pregroup grammar representation, which is finally translated into parametrized quantum circuits. This encoding mechanism follows the idea presented in QNLP \cite{Lorenz_Pearson_Meichanetzidis_Kartsaklis_Coecke_2021}. Formally, the encoding mechanism is a sequence of functor applications (Def.~\ref{def:functor}) to the abstract syntax trees of queries modeled as monoidal categories (Def.~\ref{def:monoidal_category}).

\subsection{Parsing SQL queries with context-free grammar}
The first step in the encoding is to parse the SQL queries with a context-free grammar. We utilize the context-free grammar developed for SQLite \cite{grammars-v4_2022}, but our framework can generally process any standard SQL query dialect. We assume that the \texttt{SELECT} clause consists of either a list of attributes or a standard SQL function as defined in the grammar. The \texttt{FROM} clause includes a list of relations. The \texttt{WHERE} clause contains an expression that consists of predicates, which are also expressions themselves. The predicates evaluate true or false and can be divided into filtering or join predicates.

Context-free grammar (CFG) \cite{Sipser_1997}, and its variations are the standard way to formalize query and programming languages, although they were first used to formalize natural languages. 
\begin{definition}{\textbf{Context-free grammar.}}\label{def:cfg}
A context-free grammar is a tuple $(V, \Sigma, R, s)$ where $V$ is a finite set of variables, $\Sigma$ is a finite set of terminals and disjoint from $V$, $R$ is a finite set of rules, and $s \in V$ is a start variable. The rules $R$ are elements of the set $V \times (V \cup \Sigma)^{*}$ where $()^{*}$ denotes the Kleene star operation.
\end{definition}

In other words, the rule set $R$ is a relation (in a mathematical sense) that describes how variables can be rewritten with sequences of variables and terminals. The set $R$ is also referred to as rewrite rules or productions of the grammar.

Context-free grammars induce a monoidal category defined in Def.~\ref{def:monoidal_category}, which we call a \textit{monoidal context-free grammar category}. Next, we describe how to construct this category. First, each variable in $V$ and terminal in $\Sigma$ is assigned a domain and codomain object (which we call types) to become morphisms in the monoidal context-free grammar category. A particular object (i.e., type) in the monoidal context-free grammar category is an identity type $I$, and every terminal in $\Sigma$ has it as a codomain. The collection of the types becomes the collection of objects in the monoidal context-free grammar category. The Kleene star operation induces the monoidal (tensor product) structure on the category.

Furthermore, every rewrite rule in $R$ becomes a morphism. The domain and codomain of a rewrite rule are defined by the fact that the rewrite rule maps a variable to a sequence of variables and terminals. For example, consider the following rule, which we call the $S$-rule,
\begin{displaymath}
S \to aSb,
\end{displaymath}
where $S$ is a variable and $a$ and $b$ are terminals. Then $S$, $a$ and $b$ have their corresponding types, for example, $S \colon \mathrm{dom}(S) \to \mathrm{cod}(S)$, $a \colon \mathrm{dom}(a) \to I$ and $b \colon \mathrm{dom}(b) \to I$. Thus, the type for the rule $S$ in the monoidal context-free grammar category is
\begin{displaymath}
\mathrm{cod}(S) \to \mathrm{dom}(a) \otimes \mathrm{dom}(S) \otimes \mathrm{dom}(b).
\end{displaymath} 

Rewriting rules are special morphisms because their codomain is a tensor product of the types, whereas variables and terminals always have a single type as domain and codomain. The start variable does not have a special role in the monoidal context-free grammar category.

The SQL parsers do not implement this type of typing, so the types, such as $\mathrm{dom}(S)$ and $\mathrm{dom}(a)$, are generated at the point when the abstract syntax tree is translated into a monoidal context-free grammar category. In practice, we have identified the types and implemented the monoidal context-free grammar category with Python. At a technical level, we parse SQL queries with ANTLR \cite{Parr_Quong_1995, Bovet_Parr_2008, Parr_Fisher_2011, Parr_2012} using the grammar definition for SQLite \cite{grammars-v4_2022}. This is integrated into the standard Python package to calculate with category theory: DisCoPy \cite{Felice_Toumi_Coecke_2021}, which we use to create, store, and manipulate the categories introduced in this work. DisCoPy is strongly connected to quantum and tensor computing, enabling easy transformation between different abstractions.

\begin{example}\label{ex:example1}
In this paper, we fix the following example query that demonstrates the encoding mechanism from SQL queries into parametrized circuits.
\begin{center}

    \begin{lstlisting}[language=SQL]
    SELECT planet_name, mass
    FROM planets  
    WHERE planet_name = 'Kepler-22b'; 
    \end{lstlisting}
\end{center}
After the example query has been parsed, the corresponding monoidal context-free grammar category represented as a diagram can be viewed in Figure \ref{fig:CFG_large_example}. The boxes are morphisms, and the wire labels are objects (i.e., types) as defined in Def.~\ref{def:category}. For example, \texttt{select\_stmt} is modeled as a morphism that has \texttt{statement} as a domain and codomain type, as we can see in the diagram.
\end{example}

\begin{figure*}
    \centering
    \input{figures/CFG_diagram_cat_example.tikz}
    \caption{Context-free grammar diagram of the query in Example \ref{ex:example1}}
    \label{fig:CFG_large_example}
\end{figure*}

\subsection{Converting context-free grammar to pregroup grammar}
In this part, we describe the rewriting process of transforming the context-free grammar categories into pregroup grammar categories. We develop this transformation so that the mapping between the categories is a functor, defined in Def.~\ref{def:functor}. The fact that the transformation is functorial ensures that the produced pregroup grammar category is also monoidal and the semantics of the SQL queries are preserved \cite{Lorenz_Pearson_Meichanetzidis_Kartsaklis_Coecke_2021}. The following definitions are based on \cite{Lambek_1999}. The technical details of an ordered monoid \cite{Coecke_2013} are not introduced here for space reasons.

\begin{definition}{\textbf{Pregroup.}}
A pregroup $P$ is an ordered monoid where every element $a \in P$ has left $a^{l}$ and right $a^{r}$ parts (adjoints) which satisfy the following properties with respect to the ordering $\leq$ of $P$:
    \begin{displaymath}
        a^{l}a \leq 1 \leq aa^{l} \quad \text{ and } \quad aa^{r} \leq 1 \leq a^{r}a.
    \end{displaymath}
\end{definition}
Let $B$ be a set. The free pregroup $P_B$ generated by $B$ is the smallest (with respect to set inclusion) pregroup which includes $B$.

\begin{definition}{\textbf{Pregroup grammar.}}\label{def:pregroup_grammar}
A pregroup grammar is a tuple $G = (B, \Sigma, \Delta, s)$, where the set $B$ is the finite set of basic types and the set $\Sigma$ is the finite set of the vocabulary. Let $P_B$ be the free pregroup generated by $B$. The set $\Delta \subset \Sigma \times P_B$ is a relation called the dictionary, and $s \in P_B$ is a designated sentence type. We require that $\Delta$ is finite. The sequence of words $w_1 \cdots w_n \in \Sigma^n$ is called \textit{grammatical} if for every $i = 1, \ldots, n$ there exists a tuple $(w_i, t_i) \in \Delta$ so that $t_1\cdots t_n \leq s$ in the pregoup $P_B$.
\end{definition}

The pregroup grammar defines a monoidal pregroup grammar category where objects are the elements of the disjoint union $B + \Sigma$. The entries in the dictionary $\Delta$ give the morphisms so that if $(w, t) \in \Delta$, the domain of morphism $(w, t)$ is the word $w$ and the codomain is $t$. This satisfies Def.~\ref{def:category} of a category.



The fact that the transformation is functorial ensures that the structure of the domain category, i.e., the context-free grammar category, is preserved and the outputted pregroup grammar category is grammatical in the sense of Def.~\ref{def:pregroup_grammar}. Since the current quantum hardware is limited and cannot simulate large circuits, we must perform well-reasoned simplifications while applying the functor between the different representations. When the simplifications are part of the functorial rewriting process, they do not modify the grammatical structure of the queries. This is a mathematically precise way to simplify query representations without losing the expressiveness of queries: the same query is represented with a different grammar using different syntactical information.



The complete functorial mapping from the context-free grammar category to the pregroup grammar category is technical, and due to space limitations, we demonstrate it with the following example. The key idea is to define a functor from the context-free grammar category (representing a SQL query) to the pregroup grammar category (representing the same SQL query with a different grammar), which rewrites the query in a more compact format.

\begin{example}
Continuing Example \ref{ex:example1}, Figure \ref{fig:pregroup1} demonstrates one of the transformations that the functor between the context-free grammar category and the pregroup grammar category performs at a lower level. We have taken a box that contains the \texttt{SELECT}-clause and the functor maps it to the corresponding pregroup grammar element, which is given by the dashed lines. We can see that the box disappears, simplifying the query representation of Fig.~\ref{fig:CFG_large_example}. The objects at the end of the wires are mapped to the abstract pregroup grammar types. Fig.~\ref{fig:pregroup_large_example1} shows the complete pregroup grammar diagram transformed from the context-free grammar diagram in Fig.~\ref{fig:CFG_large_example}. As we can see, most of the boxes have disappeared, but we can still read the original query with the additional structure information that the bent wires bring. These wires describe the relationships between the elements from the context-free grammar representation. The mappings are manually designed, but the application of the mappings is automated with DisCoPy.
\end{example}

\begin{figure*}
  \begin{subfigure}[t]{0.33\textwidth}
    \centering
    \begin{adjustbox}{width=0.8\textwidth}
      \input{figures/Nora_presentation_diagram_CFG_to_pregroup_1.tikz}
    \end{adjustbox}
    \caption{A part of the functorial transformation}
    \label{fig:pregroup1}
  \end{subfigure}
  \hspace{0.5em}
  \rulesep
  \hspace{0.5em}
  \begin{subfigure}[t]{0.3\textwidth}
    \centering
    \begin{adjustbox}{width=0.8\textwidth}
      \input{figures/Nora_presentation_diagram_CFG_to_pregroup_3.tikz}
    \end{adjustbox}
    \caption{One of the functorial transformations removes caps from the pregroup grammar diagrams}
      \label{fig:pregroup2}
  \end{subfigure}
  \hspace{0.5em}
  \rulesep
  \hspace{0.5em}
  \begin{subfigure}[t]{0.25\textwidth}
    \centering
    \begin{adjustbox}{width=1\textwidth}
      \input{figures/cupless_pregroup_diagram_cat_example.tikz}
    \end{adjustbox}
    \caption{Pregroup grammar diagram of the query in Example \ref{ex:example1} without caps}
      \label{fig:cupless_pregroup_diagram_cat}
  \end{subfigure}
  \caption{Functorial transformations to the context-free and pregroup grammar diagrams}
\end{figure*}

\begin{figure}
    \centering
    \input{figures/pregroup_diagram_cat_example.tikz}
    \caption{The pregroup grammar diagram of the query in Example \ref{ex:example1}.}
    \label{fig:pregroup_large_example1}
\end{figure}

Before constructing the circuits, we perform a simplification process to remove the caps from the pregroup grammar diagrams. For example, in Fig.~\ref{fig:pregroup_large_example1}, we can see eight caps, which are the bent wires between the boxes. Caps denote the grammar reductions, and removing them is a standard rewriting technique \cite{Lorenz_Pearson_Meichanetzidis_Kartsaklis_Coecke_2021}. Pregroup diagrams with no caps lead to circuits with fewer qubits and less need for costly post-processing. Besides, implementing a cap requires obtaining expensive quantum measurement results that happen non-deterministically \cite{Lorenz_Pearson_Meichanetzidis_Kartsaklis_Coecke_2021}. In our work, the cap removal procedure is a functor between two pregroup grammar diagrams. See Fig.~\ref{fig:pregroup2}, which demonstrates one of the cap removal rewriting rules: we can intuitively think that we bend the wires and make them straight.

\subsection{Converting pregroup grammar to circuits}
In the final phase of the SQL query encoding mechanism, we translate the produced pregroup grammar categories into parametrized quantum circuits. These circuits form the core of our quantum machine learning model. To perform the last translation, we use Lambeq \cite{kartsaklis2021lambeq} software package, which is a modular and extensible high-level Python library for experimental quantum natural language processing. 

The Lambeq framework implements various functors that translate pregroup grammar categories into parametrized quantum circuits. This mapping operates at the structural level, where it maps boxes in the pregroup grammar diagram to gate combinations in the circuit. Due to space limits, we skip the technical details of these functors, and they are relatively well developed in QNLP \cite{kartsaklis2021lambeq}. The high-level idea is to map the wires in pregroup grammar diagrams into wires in quantum circuits and boxes into collections of parametrized quantum gates. This way, we will have a one-to-one correspondence between the SQL query elements and sets of parameters in the parametrized circuit. For instance, a keyword or attribute in the query corresponds to a particular set of parameters in the circuit.

Some of the functors in Lambeq are IQPAnsatz \cite{Shepherd_2009, Havl_ek_2019} and Sim14Ansatz, which refers to the ansatz number 14 in \cite{Sim_Johnson_Aspuru_Guzik_2019}. The functors differ depending on which quantum gate layouts are mapped to the boxes in the pregroup grammar diagram. We evaluated IQPAnsatz and Sim14Ansatz and deduced substantial performance differences between these layouts, which we will discuss in the results section.

\begin{example}
See Fig.~\ref{fig:cupless_pregroup_diagram_cat} as an example of a capless pregroup grammar diagram based on Example \ref{ex:example1}. Fig.~\ref{fig:cat_circuit} illustrates the final circuit with randomly substituted variables after we have applied the IQPAnsatz functor in Lambeq. This representation finalizes the query encoding example, which we started in Example \ref{ex:example1}. Other queries produce circuits with different structures. This sequence of examples demonstrates how the example SQL query can be translated into a parametrized quantum circuit with multiple functorial transformation steps.
\end{example}



\begin{figure*}
    \centering
    \input{figures/cat_circuit.tex}
    \caption{The circuit corresponding the query in Example \ref{ex:example1} with random parameter values}
    \label{fig:cat_circuit}
\end{figure*}


%% file: figures/CFG_diagram_cat_example.tikz
\resizebox{0.8\textwidth}{!}{%
\begin{tikzpicture}
	\begin{pgfonlayer}{nodelayer}
		\node [style=none] (0) at (-0.1, 3.5) {};
		\node [style=none] (1) at (-0.1, 3) {};
		\node [style=none] (2) at (-0.1, 3) {};
		\node [style=none] (3) at (-0.1, 2.25) {};
		\node [style=none, fill=white, right] (4) at (0.25, 2.65) {list};
		\node [style=none] (5) at (-0.1, 2.25) {};
		\node [style=none] (6) at (-0.1, 1.5) {};
		\node [style=none, fill=white, right] (7) at (0.5, 1.9) {statement};
		\node [style=none] (8) at (-0.1, 1.5) {};
		\node [style=none] (9) at (-0.1, 0.75) {};
		\node [style=none, fill=white, right] (10) at (0.5, 1.15) {statement};
		\node [style=none] (11) at (-0.1, 0.75) {};
		\node [style=none] (12) at (-0.1, 0) {};
		\node [style=none, fill=white, right] (13) at (0.5, 0.4) {statement};
		\node [style=none, fill=white, right] (16) at (-7.5, 1.4) {select\_clause};
		\node [style=none] (17) at (-0.1, 0) {};
		\node [style=none] (18) at (-0.1, -1.5) {};
		\node [style=none, fill=white, right] (19) at (0, -1.1) {from\_clause};
		\node [style=none] (20) at (0.4, 0) {};
		\node [style=none] (21) at (6.65, 0.5) {};
		\node [style=none, fill=white, right] (22) at (6, 1.4) {where\_clause};
		\node [style=none] (44) at (-0.6, -1.5) {};
		\node [style=none] (45) at (-0.6, -2.25) {};
		\node [style=none, fill=white, right] (46) at (-3.5, -2.1) {from\_keyword};
		\node [style=none] (47) at (0.4, -1.5) {};
		\node [style=none] (48) at (0.4, -2.25) {};
		\node [style=none, fill=white, right] (49) at (0.75, -1.85) {table};
		\node [style=none] (50) at (0.4, -2.25) {};
		\node [style=none] (51) at (0.4, -3) {};
		\node [style=none, fill=white, right] (52) at (0.75, -2.6) {table\_name};
		\node [style=none] (53) at (6.4, 0.5) {};
		\node [style=none] (54) at (5.65, -1) {};
		\node [style=none, fill=white, right] (55) at (3.5, -0.6) {where\_keyword};
		\node [style=none] (56) at (6.9, 0.5) {};
		\node [style=none] (57) at (7.9, -1) {};
		\node [style=none, fill=white, right] (58) at (8, -0.6) {expr};
		\node [style=none] (59) at (7.65, -1) {};
		\node [style=none] (60) at (6.15, -2.25) {};
		\node [style=none, fill=white, right] (61) at (4, -1.85) {binary\_operator};
		\node [style=none] (62) at (7.9, -1) {};
		\node [style=none] (63) at (7.9, -2.25) {};
		\node [style=none, fill=white, right] (64) at (8, -1.85) {column\_expr};
		\node [style=none] (65) at (8.15, -1) {};
		\node [style=none] (66) at (10.15, -2.25) {};
		\node [style=none, fill=white, right] (67) at (10.25, -1.85) {expr};
		\node [style=none] (68) at (7.9, -2.25) {};
		\node [style=none] (69) at (7.9, -3.25) {};
		\node [style=none, fill=white, right] (70) at (8, -2.85) {column\_name};
		\node [style=none] (71) at (10.15, -2.25) {};
		\node [style=none] (72) at (10.15, -3.25) {};
		\node [style=none, fill=white, right] (73) at (10.25, -2.85) {literal\_value};
		\node [style=box] (75) at (-0.1, 3) {parse};
		\node [style=none] (76) at (0.4, 3.5) {query};
		\node [style=box] (77) at (10.15, -3.25) {'Kepler-22b'};
		\node [style=box] (78) at (10.15, -2.25) {literal-expr};
		\node [style=box] (79) at (7.9, -3.25) {planet\_name};
		\node [style=box] (80) at (7.9, -2.25) {column-expr};
		\node [style=box] (81) at (6.15, -2.25) {=};
		\node [style=box] (82) at (7.9, -1) {bin-expr};
		\node [style=box] (83) at (5.65, -1) {WHERE};
		\node [style=box] (84) at (6.65, 0.5) {where-clause};
		\node [style=box] (86) at (-0.1, 2.25) {sql\_stmt\_list};
		\node [style=box] (87) at (-0.1, 1.5) {sql\_stmt};
		\node [style=box] (88) at (-0.1, 0.75) {select\_stmt};
		\node [style=box] (89) at (-0.1, 0) {select-core};
		\node [style=box] (96) at (-0.1, -1.5) {from-clause};
		\node [style=box] (97) at (0.4, -2.25) {table};
		\node [style=box] (98) at (0.4, -3) {planets};
		\node [style=box] (99) at (-0.6, -2.25) {FROM};
		\node [style=box] (101) at (-7.75, 0.5) {select-clause};
		\node [style=box] (102) at (-10.75, -1.25) {SELECT};
		\node [style=box] (103) at (-7.75, -1.25) {result-column};
		\node [style=box] (104) at (-5.25, -1.25) {result-column};
		\node [style=box] (105) at (-7.75, -2.25) {column-expr};
		\node [style=box] (106) at (-7.75, -3.25) {planet\_name};
		\node [style=box] (107) at (-5.25, -2.25) {column-expr};
		\node [style=box] (108) at (-5.25, -3.25) {mass};
		\node [style=none] (109) at (-0.5, 0) {};
		\node [style=none] (111) at (-9.25, -0.75) {select\_keyword};
		\node [style=none] (112) at (-6.75, -0.75) {result\_column};
		\node [style=none] (113) at (-6.75, -1.75) {column\_expr};
		\node [style=none] (114) at (-6.75, -2.75) {column\_name};
		\node [style=none] (115) at (-4.25, -0.75) {result\_column};
		\node [style=none] (116) at (-4.25, -1.75) {column\_expr};
		\node [style=none] (117) at (-4.25, -2.75) {column\_name};
		\node [style=none] (118) at (-8.25, 0.5) {};
		\node [style=none] (119) at (-7.25, 0.5) {};
	\end{pgfonlayer}
	\begin{pgfonlayer}{edgelayer}
		\draw [in=90, out=-90] (0.center) to (1.center);
		\draw [in=90, out=-90] (2.center) to (3.center);
		\draw [in=90, out=-90] (5.center) to (6.center);
		\draw [in=90, out=-90] (8.center) to (9.center);
		\draw [in=90, out=-90] (11.center) to (12.center);
		\draw [in=90, out=-90] (17.center) to (18.center);
		\draw [in=90, out=-90, looseness=0.50] (20.center) to (21.center);
		\draw [in=90, out=-90] (44.center) to (45.center);
		\draw [in=90, out=-90] (47.center) to (48.center);
		\draw [in=90, out=-90] (50.center) to (51.center);
		\draw [in=90, out=-90] (53.center) to (54.center);
		\draw [in=90, out=-90] (56.center) to (57.center);
		\draw [in=90, out=-90] (59.center) to (60.center);
		\draw [in=90, out=-90] (62.center) to (63.center);
		\draw [in=90, out=-90] (65.center) to (66.center);
		\draw [in=90, out=-90] (68.center) to (69.center);
		\draw [in=90, out=-90] (71.center) to (72.center);
		\draw (103) to (101);
		\draw (103) to (105);
		\draw (104) to (107);
		\draw (105) to (106);
		\draw (107) to (108);
		\draw [in=-90, out=90, looseness=0.50] (101) to (109.center);
		\draw [in=-90, out=90, looseness=0.75] (102) to (118.center);
		\draw [in=90, out=-90, looseness=0.50] (119.center) to (104);
	\end{pgfonlayer}
\end{tikzpicture}

}

%% file: figures/Nora_presentation_diagram_CFG_to_pregroup_1.tikz
\begin{tikzpicture}
	\begin{pgfonlayer}{nodelayer}
		\node [style=none] (1) at (-15, 0.5) {};
		\node [style=none] (4) at (-12.25, 0.5) {};
		\node [style=none] (6) at (-15, -0.25) {};
		\node [style=none] (7) at (-15.75, 1.5) {};
		\node [style=none] (8) at (-11.5, 1.5) {};
		\node [style=none] (9) at (-16.5, -0.25) {};
		\node [style=none] (10) at (-16.5, 3) {};
		\node [style=none] (16) at (-17, 2.5) {};
		\node [style=none] (17) at (-17, 0.75) {};
		\node [style=none] (18) at (-10.75, 2.5) {};
		\node [style=none] (19) at (-10.75, 0.75) {};
		\node [style=none] (20) at (-15.75, -0.25) {};
		\node [style=none] (21) at (-12.25, -1.5) {};
		\node [style=none] (22) at (-11.5, -2) {};
		\node [style=none] (23) at (-17, -0.25) {};
		\node [style=none] (24) at (-14.5, -0.25) {};
		\node [style=none] (25) at (-15.75, -0.75) {};
		\node [style=none] (26) at (-14.75, -1.25) {select\_keyword $\mapsto$ n $\otimes$ n.l $\otimes$ n.l};
		\node [style=none] (27) at (-15, 3) {select\_clause $\mapsto$ n};
		\node [style=none] (30) at (-14, 3.75) {$\Downarrow$};
		\node [style=box] (31) at (-14, 6) {select-clause};
		\node [style=none] (39) at (-14, 6.75) {};
		\node [style=none] (40) at (-13, 6.5) {select\_clause};
		\node [style=none] (41) at (-16.5, 4.5) {select\_keyword};
		\node [style=none] (42) at (-14, 4.5) {result\_column};
		\node [style=none] (45) at (-11.5, 4.5) {result\_column};
		\node [style=none] (48) at (-14.5, 6) {};
		\node [style=none] (50) at (-13.5, 6) {};
		\node [style=none] (51) at (-16.5, 4.75) {};
		\node [style=none] (52) at (-14, 4.75) {};
		\node [style=none] (53) at (-11.5, 4.75) {};
		\node [style=none] (54) at (-16.25, 0.25) {n};
		\node [style=none] (55) at (-14.75, 0.25) {n.l};
		\node [style=none] (56) at (-15.5, 0.25) {n.l};
		\node [style=none] (57) at (-12, 0) {n};
		\node [style=none] (58) at (-11.25, 0) {n};
		\node [style=none] (59) at (-13, -1.75) {result\_column $\mapsto$ n};
		\node [style=none] (61) at (-12, -2.25) {result\_column $\mapsto$ n};
		\node [style=none] (62) at (-15.75, -1) {};
	\end{pgfonlayer}
	\begin{pgfonlayer}{edgelayer}
		\draw [bend left=90] (1.center) to (4.center);
		\draw (6.center) to (1.center);
		\draw [bend left=90, looseness=0.50] (7.center) to (8.center);
		\draw (9.center) to (10.center);
		\draw [style=dashed edge] (18.center) to (19.center);
		\draw [style=dashed edge] (19.center) to (17.center);
		\draw [style=dashed edge] (17.center) to (16.center);
		\draw [style=dashed edge] (16.center) to (18.center);
		\draw (7.center) to (20.center);
		\draw (4.center) to (21.center);
		\draw (8.center) to (22.center);
		\draw [in=135, out=-90, looseness=0.75] (23.center) to (25.center);
		\draw [in=-90, out=30, looseness=0.75] (25.center) to (24.center);
		\draw (31) to (39.center);
		\draw [in=-90, out=90, looseness=0.50] (51.center) to (48.center);
		\draw (31) to (52.center);
		\draw [in=90, out=-90, looseness=0.50] (50.center) to (53.center);
		\draw (25.center) to (62.center);
	\end{pgfonlayer}
\end{tikzpicture}%

%% file: figures/Nora_presentation_diagram_CFG_to_pregroup_3.tikz
\begin{tikzpicture}
	\begin{pgfonlayer}{nodelayer}
		\node [style=box] (0) at (-2, -1.25) {SELECT};
		\node [style=none] (1) at (-1.75, -0.25) {};
		\node [style=box] (2) at (0, -1.25) {planet\_name};
		\node [style=box] (3) at (1.75, -1.25) {mass};
		\node [style=none] (4) at (0, -0.25) {};
		\node [style=none] (5) at (-2, -1.25) {};
		\node [style=none] (6) at (-1.75, -1.25) {};
		\node [style=none] (7) at (-2, 0.75) {};
		\node [style=none] (8) at (1.75, 0.75) {};
		\node [style=none] (9) at (-2.25, -1.25) {};
		\node [style=none] (10) at (-2.25, 1.75) {};
		\node [style=none] (11) at (-2.5, -0.75) {n};
		\node [style=none] (12) at (-1.5, 0.75) {n.l};
		\node [style=none] (13) at (-1.25, -0.25) {n.l};
		\node [style=none] (14) at (0.25, -0.25) {n};
		\node [style=none] (15) at (2, -0.25) {n};
		\node [style=box] (16) at (-0.25, -5.25) {SELECT};
		\node [style=none] (17) at (2, -3.25) {};
		\node [style=box] (18) at (2, -3.25) {planet\_name};
		\node [style=box] (19) at (-0.25, -3.25) {mass};
		\node [style=none] (21) at (-0.25, -5.25) {};
		\node [style=none] (22) at (0, -5.25) {};
		\node [style=none] (23) at (-0.25, -3.25) {};
		\node [style=none] (25) at (-0.5, -5.25) {};
		\node [style=none] (26) at (-2.25, -3.25) {};
		\node [style=none] (27) at (-1.5, -4.5) {n};
		\node [style=none] (28) at (0, -4.5) {n.l};
		\node [style=none] (29) at (1.25, -4.5) {n.l};
		\node [style=none] (31) at (-0.25, -2.25) {$\Downarrow$};
	\end{pgfonlayer}
	\begin{pgfonlayer}{edgelayer}
		\draw [bend left=90] (1.center) to (4.center);
		\draw (4.center) to (2);
		\draw (6.center) to (1.center);
		\draw (5.center) to (7.center);
		\draw [bend left=90, looseness=0.50] (7.center) to (8.center);
		\draw (8.center) to (3);
		\draw (9.center) to (10.center);
		\draw [in=-90, out=90, looseness=0.50] (22.center) to (17.center);
		\draw (21.center) to (23.center);
		\draw [in=-90, out=90] (25.center) to (26.center);
	\end{pgfonlayer}
\end{tikzpicture}

%% file: figures/cupless_pregroup_diagram_cat_example.tikz
\begin{tikzpicture}
	\begin{pgfonlayer}{nodelayer}
		\node [style=none] (1) at (-1.25, 2.75) {};
		\node [style=none] (2) at (-1.25, 0.75) {};
		\node [style=none, fill=white, right] (3) at (-1.65, 2.25) {s};
		\node [style=none] (4) at (-0.25, 3.5) {};
		\node [style=none] (5) at (-0.25, 1.75) {};
		\node [style=none, fill=white, right] (6) at (-0.65, 2.65) {n.l};
		\node [style=none] (7) at (1, 2.75) {};
		\node [style=none] (8) at (1, 1.75) {};
		\node [style=none, fill=white, right] (9) at (1.1, 2.15) {n.l};
		\node [style=none] (10) at (0.25, 1.75) {};
		\node [style=none] (11) at (0.25, 0.75) {};
		\node [style=none, fill=white, right] (12) at (0.6, 1.15) {n.l};
		\node [style=none] (13) at (-0.5, 0.75) {};
		\node [style=none] (14) at (-0.5, -0.75) {};
		\node [style=none, fill=white, right] (15) at (-0.9, -0.1) {n};
		\node [style=none] (16) at (0.5, 0) {};
		\node [style=none] (17) at (0.5, -0.75) {};
		\node [style=none, fill=white, right] (18) at (0.85, -0.35) {n.l};
		\node [style=none] (19) at (0, -0.75) {};
		\node [style=none] (20) at (0, -3) {};
		\node [style=none, fill=white, right] (21) at (-0.4, -2.1) {n};
		\node [style=none] (22) at (1.25, -1.5) {};
		\node [style=none] (23) at (1.25, -3) {};
		\node [style=none, fill=white, right] (24) at (0.85, -2.1) {n.l};
		\node [style=none] (25) at (2.5, -2.25) {};
		\node [style=none] (26) at (2.5, -3) {};
		\node [style=none, fill=white, right] (27) at (2.85, -2.6) {n.l};
		\node [style=box] (73) at (-0.25, 3.5) {'Kepler-22b'};
		\node [style=box] (74) at (1, 2.75) {planet\_name};
		\node [style=box] (75) at (0.25, 1.75) {$ \quad $ = $ \quad \ $};
		\node [style=box] (76) at (-0.5, 0.75) {$\ $ WHERE $\ $};
		\node [style=box] (77) at (0.5, 0) {planets};
		\node [style=box] (78) at (0, -0.75) {$\ $ FROM $\ $};
		\node [style=box] (79) at (1.25, -1.5) {mass};
		\node [style=box] (80) at (2.5, -2.25) {planet\_name};
		\node [style=box] (81) at (1.25, -3) {$\quad \quad $ SELECT $\quad \quad $};
	\end{pgfonlayer}
	\begin{pgfonlayer}{edgelayer}
		\draw [in=90, out=-90] (1.center) to (2.center);
		\draw [in=90, out=-90] (4.center) to (5.center);
		\draw [in=90, out=-90] (7.center) to (8.center);
		\draw [in=90, out=-90] (10.center) to (11.center);
		\draw [in=90, out=-90] (13.center) to (14.center);
		\draw [in=90, out=-90] (16.center) to (17.center);
		\draw [in=90, out=-90] (19.center) to (20.center);
		\draw [in=90, out=-90] (22.center) to (23.center);
		\draw [in=90, out=-90] (25.center) to (26.center);
	\end{pgfonlayer}
\end{tikzpicture}

%% file: figures/pregroup_diagram_cat_example.tikz
\resizebox{\linewidth}{!}{%
\begin{tikzpicture}
	\begin{pgfonlayer}{nodelayer}
		\node [style=none] (0) at (3.5, 2.25) {};
		\node [style=none] (1) at (3.5, -1.25) {};
		\node [style=none, fill=white, right] (2) at (3.6, 1.75) {s};
		\node [style=none] (4) at (-4.25, 1) {};
		\node [style=none] (5) at (0, 1) {};
		\node [style=none] (6) at (-4.25, -0.5) {};
		\node [style=none, fill=white, right] (7) at (-3.9, 1.4) {n};
		\node [style=none] (8) at (0, -1.25) {};
		\node [style=none, fill=white, right] (9) at (-0.15, 0.9) {n.r};
		\node [style=none] (11) at (0.25, 1.5) {};
		\node [style=none] (12) at (3.25, 1.5) {};
		\node [style=none] (13) at (0.25, -1.25) {};
		\node [style=none, fill=white, right] (14) at (0.6, 1.4) {n};
		\node [style=none] (15) at (3.25, -1.25) {};
		\node [style=none, fill=white, right] (16) at (2.85, 1.4) {n.r};
		\node [style=none] (18) at (-4, 0.5) {};
		\node [style=none] (19) at (-1, 0.5) {};
		\node [style=none] (20) at (-4, -0.5) {};
		\node [style=none, fill=white, right] (21) at (-3.65, 0.9) {n.l};
		\node [style=none] (22) at (-1, -0.5) {};
		\node [style=none, fill=white, right] (23) at (-0.65, 0.65) {n};
		\node [style=none] (25) at (-3.75, 0) {};
		\node [style=none] (26) at (-2.5, 0) {};
		\node [style=none] (27) at (-3.75, -0.5) {};
		\node [style=none, fill=white, right] (28) at (-3.4, 0.4) {n.l};
		\node [style=none] (29) at (-2.5, -1.25) {};
		\node [style=none, fill=white, right] (30) at (-2.4, 0.15) {n};
		\node [style=none] (32) at (0.5, 0.75) {};
		\node [style=none] (33) at (1.75, 0.75) {};
		\node [style=none] (34) at (0.5, -1.25) {};
		\node [style=none, fill=white, right] (35) at (0.85, 0.4) {n.l};
		\node [style=none] (36) at (1.75, -0.5) {};
		\node [style=none, fill=white, right] (37) at (2.1, 0.4) {n};
		\node [style=none] (39) at (3.75, 0.5) {};
		\node [style=none] (40) at (4.5, 0.5) {};
		\node [style=none] (41) at (3.75, -1.25) {};
		\node [style=none, fill=white, right] (42) at (3.85, 0.15) {n.l};
		\node [style=none] (43) at (4.5, -0.5) {};
		\node [style=none, fill=white, right] (44) at (4.35, 0.9) {n};
		\node [style=none] (46) at (4.75, 0.25) {};
		\node [style=none] (47) at (7.75, 0.25) {};
		\node [style=none] (48) at (4.75, -0.5) {};
		\node [style=none, fill=white, right] (49) at (5.1, 0.65) {n.l};
		\node [style=none, fill=white, right] (51) at (7.85, 0.4) {n};
		\node [style=none] (53) at (5, 0) {};
		\node [style=none] (54) at (6, 0) {};
		\node [style=none] (55) at (5, -0.5) {};
		\node [style=none, fill=white, right] (56) at (5.35, 0.15) {n.l};
		\node [style=none] (57) at (6, -1.25) {};
		\node [style=none, fill=white, right] (58) at (6.1, 0.15) {n};
		\node [style=box] (99) at (7.75, -0.5) {'Kepler-22b'};
		\node [style=box] (100) at (6, -1.25) {planet\_name};
		\node [style=box] (101) at (4.75, -0.5) {$\ $ = $\ $};
		\node [style=box] (102) at (-1, -0.5) {mass};
		\node [style=box] (103) at (-2.5, -1.25) {planet\_name};
		\node [style=box] (104) at (-4, -0.5) {SELECT};
		\node [style=box] (105) at (0.25, -1.25) {FROM};
		\node [style=box] (106) at (1.75, -0.5) {planets};
		\node [style=box] (107) at (3.5, -1.25) {WHERE};
	\end{pgfonlayer}
	\begin{pgfonlayer}{edgelayer}
		\draw [in=90, out=-90] (0.center) to (1.center);
		\draw [in=90, out=-90] (4.center) to (6.center);
		\draw [in=90, out=-90] (5.center) to (8.center);
		\draw [in=90, out=-90] (11.center) to (13.center);
		\draw [in=90, out=-90] (12.center) to (15.center);
		\draw [in=90, out=-90] (18.center) to (20.center);
		\draw [in=90, out=-90] (19.center) to (22.center);
		\draw [in=90, out=-90] (25.center) to (27.center);
		\draw [in=90, out=-90] (26.center) to (29.center);
		\draw [in=90, out=-90] (32.center) to (34.center);
		\draw [in=90, out=-90] (33.center) to (36.center);
		\draw [in=90, out=-90] (39.center) to (41.center);
		\draw [in=90, out=-90] (40.center) to (43.center);
		\draw [in=90, out=-90] (46.center) to (48.center);
		\draw [in=90, out=-90] (53.center) to (55.center);
		\draw [in=90, out=-90] (54.center) to (57.center);
		\draw (47.center) to (99);
		\draw [bend left=90] (46.center) to (47.center);
		\draw [bend left=90, looseness=1.50] (53.center) to (54.center);
		\draw [bend left=90] (4.center) to (5.center);
		\draw [bend left=90] (18.center) to (19.center);
		\draw [bend left=90, looseness=2.25] (25.center) to (26.center);
		\draw [bend left=90, looseness=1.25] (32.center) to (33.center);
		\draw [bend left=90] (11.center) to (12.center);
		\draw [bend left=90, looseness=1.75] (39.center) to (40.center);
	\end{pgfonlayer}
\end{tikzpicture}%
}

%% file: figures/cat_circuit.tex
\resizebox{0.9\textwidth}{!}{
\Qcircuit @C=0.5em @R=0.2em @!R { \\
	 	\nghost{{q}_{0} : } & \lstick{{q}_{0} : {\ket{0}} } & \gate{\mathrm{H}} & \qw & \ctrl{1} & \gate{\mathrm{H}} & \ctrl{3} & \gate{\mathrm{R_X}\,(\mathrm{0.5257})} & \gate{\mathrm{H}} & \ctrl{4} & \gate{\mathrm{R_X}\,(\mathrm{0.6727})} & \qw & \qw & \qw & \qw & \qw & \qw & \qw & \qw \\
	 	\nghost{{q}_{1} :  } & \lstick{{q}_{1} : {\ket{0}} } & \gate{\mathrm{H}} & \ctrl{1} & \gate{\mathrm{R_Z}\,(\mathrm{4.471})} & \gate{\mathrm{R_X}\,(\mathrm{4.255})} & \qw & \qw & \qw & \qw & \qw & \qw & \qw & \qw & \qw & \qw & \qw & \ket{0} & \\
	 	\nghost{{q}_{2} :  } & \lstick{{q}_{2} : {\ket{0}} } & \gate{\mathrm{H}} & \gate{\mathrm{R_Z}\,(\mathrm{0.2589})} & \gate{\mathrm{R_X}\,(\mathrm{1.956})} & \qw & \qw & \qw & \qw & \qw & \qw & \qw & \qw & \qw & \qw & \qw & \qw & \ket{0} & \rstick{\quad\text{post-}} \\
	 	\nghost{{q}_{3} :  } & \lstick{{q}_{3} : {\ket{0}} } & \gate{\mathrm{H}} & \qw & \qw & \qw & \gate{\mathrm{R_Z}\,(\mathrm{0.5257})} & \gate{\mathrm{R_X}\,(\mathrm{1.331})} & \gate{\mathrm{R_X}\,(\mathrm{4.703})} & \qw & \qw & \qw & \qw & \qw & \qw & \qw & \qw & \ket{0} & \rstick{\quad\text{se-}} \\
	 	\nghost{{q}_{4} :  } & \lstick{{q}_{4} : {\ket{0}} } & \gate{\mathrm{H}} & \qw & \qw & \qw & \qw & \qw & \qw & \gate{\mathrm{R_Z}\,(\mathrm{0.6727})} & \gate{\mathrm{R_X}\,(\mathrm{3.261})} & \gate{\mathrm{H}} & \ctrl{1} & \gate{\mathrm{R_X}\,(\mathrm{1.042})} & \gate{\mathrm{R_X}\,(\mathrm{4.272})} & \qw & \qw & \ket{0} & \rstick{\quad\text{lec-}} \\
	 	\nghost{{q}_{5} :  } & \lstick{{q}_{5} : {\ket{0}} } & \gate{\mathrm{H}} & \qw & \qw & \qw & \qw & \qw & \qw & \qw & \qw & \qw & \gate{\mathrm{R_Z}\,(\mathrm{1.042})} & \gate{\mathrm{R_X}\,(\mathrm{0.4325})} & \gate{\mathrm{R_X}\,(\mathrm{1.956})} & \qw & \qw & \ket{0} & \rstick{\quad\text{tion}} {\gategroup{3}{19}{7}{19}{1em}{\}}} \\
   }}

%% file: sections/training_main.tex
\section{Training pipeline}\label{sec:training}
This section presents the training pipeline we have used to train the proposed quantum machine learning model. 

\subsection{Data preprocessing}

\textbf{Data collection and splitting.} We have created a data set of queries to train, validate, and test the developed quantum machine learning model. The synthetic dataset has been used because it is necessary to have control over the SQL query length: the previously described encoding mechanism shows that the length of the query corresponds with how many qubits have to be used. The queries must be short since the current simulations are limited to around 20 qubits \cite{bowles2024betterclassicalsubtleart}. Although the queries are short, the experimental evaluation shows that making estimations for them classically is not necessarily easier. 

The set of synthetic queries is based on the Join Order Benchmark \cite{Leis_Gubichev_Mirchev_Boncz_Kemper_Neumann_2015} and Internet Movie Database \cite{IMDb}. We have decomposed the JOB benchmark queries into two documents, which work as seeds for generating synthetic queries. We executed the generated queries using PostgreSQL 14.2 
The execution time, cost, and cardinality statistics are gathered from PostgreSQL using the \texttt{EXPLAIN} \texttt{ANALYZE} command. The queries were randomly split into training, validation, and test queries so that all classification tasks contained around 400 training queries, 135 validation queries, and 140 test queries.

As we described in the beginning, we are solving a classification task. We have three classification tasks for every metric: binary, 4-class, and 8-class. Thus, depending on the classification task, the queries are divided into two, four, or eight equally sized sets for training and evaluation. This process defines the classes, and they are concretely listed in Table~\ref{tab:statistics2}. Compared to most previous quantum machine learning models \cite{bowles2024betterclassicalsubtleart}, implementing 4-class or 8-class classifications in quantum machine learning is a novel and relatively rare approach.

\input{plots/circuit_statistics_classes.tex}


\textbf{Encoding.} The generated queries are encoded as parametrized circuits following the encoding mechanism described in Section \ref{sec:sqltocirc}. The procedure enables us to precisely map a query and its corresponding encoded, parametrized circuit, ensuring that every query is associated with a unique quantum circuit. Furthermore, due to functorial transformations, the query's structure corresponds to the circuit's structure. To summarize this connection, every SQL keyword, identifier, literal, function, and operator is mapped to a combination of parametrized gates. The number of keywords and operators in the query gives the connectivity between the qubits. Although this process may be theoretically complex, it is rule-based, does not involve computationally complex elements, and operates on queries whose length is always limited. Thus, we believe it can be implemented efficiently even for large queries.





\subsection{Training}

\textbf{Quantum subroutine.} Quantum machine learning models usually consist of quantum and classical subroutines. Our model admits a quantum subroutine responsible for executing the parametrized circuits on a simulator. In the first epoch of the training, the parameters are initialized randomly. Our implementation and simulator are based on Pennylane \cite{pennylane}. 

Before interpreting the classification result from the measurement, the model requires performing a post-selection \cite{Aaronson_2005}. This requirement follows from the encoding mechanism that maps SQL queries into circuits. Post-selection refers to conditioning the outcome of the target qubits based on the measured results of the non-target qubits. Fig.~\ref{fig:cat_circuit} shows an example of how post-selection is performed in the case of binary classification. The target qubit is the top qubit in the figure, and the rest are non-target qubits. The non-target qubits must be in the $0$-state after the measurement. If they all are in a $0$ state, the result ($0$ or $1$) from the target qubit is accepted. Otherwise, the measurement result is discarded. If we measure a $0$, the model outputs that the SQL query corresponding to the executed circuit belongs to class $0$, which corresponds to the range used for estimation and training.

\textbf{Classical subroutine.} Besides the quantum subroutine, the model admits a classical subroutine responsible for optimizing the parameters in the quantum circuits using classical machine learning optimizers. The classical optimizer estimates the gradient based on the measured results from the quantum circuits. We have used Adam with weight decay from Optax \cite{deepmind2020jax}. The loss function is cross-entropy.

\begin{figure*}
    \centering
    \includegraphics[width=0.7\linewidth]{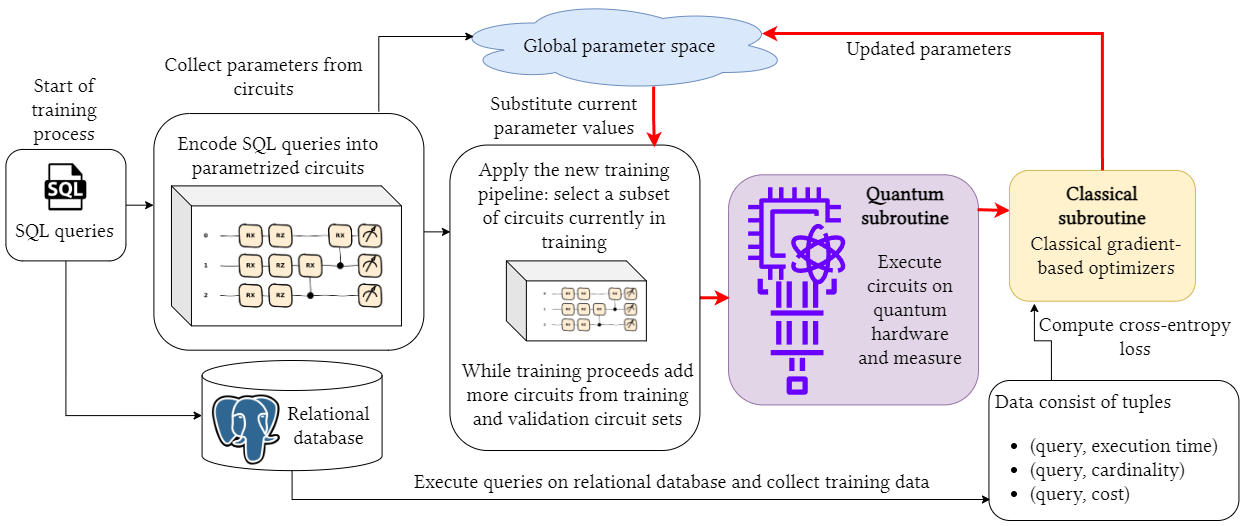}
    \caption{Training pipeline in the proposed quantum machine learning model with the novel incremental method. A single training loop indicated with the red arrows is executed for each set of circuits, whose number is incrementally increased.}
    \label{fig:optimization_workflow}
\end{figure*}

In the entire training pipeline, the quantum and classical subroutines are repeated as represented in Fig.~\ref{fig:optimization_workflow}. In contrast to the previous quantum machine learning models, the model implements an iterative training pipeline to address some challenges that quantum machine learning faces \cite{bowles2024betterclassicalsubtleart}. We begin with a relatively small set of parametrized quantum circuits (20 circuits). After applying the training pipeline to this initial set, we expand the pool by adding 20 new circuits and repeating the training process. We repeat the process until we have added all the training circuits. This approach yielded better results than training all 400 circuits simultaneously. While this method is simple, we suggest it could be used in many other similar quantum machine learning setups where the model consists of multiple circuits.

%% file: plots/circuit_statistics_classes.tex

\begin{table*}[t]
        \centering
        \resizebox{0.99\linewidth}{!}{%
    \begin{tabular}{|c|c|c|c|}
    \hline
    Task & \#classes & Classes for training, validation and test & Results \\
    \hline
    execution time & 2 & (0, 264) (264, 15603) ms & Fig. \ref{fig:9_pennylane_optax_adamw_state_IQPAnsatz_execution_time_20_20_001_2} \\
    \hline
    cardinality & 2 & (0, 33181) (33182, 102540525) rows & Fig. \ref{fig:1_pennylane_optax_adamw_state_IQPAnsatz_cardinality_20_20_001_2} \\
    \hline
    cost & 2 & (0, 83723) (83723, 1425540) psql cost & Fig. \ref{fig:5_pennylane_optax_adamw_state_IQPAnsatz_cost_20_20_001_2} \\
    \hline
    \hline
    execution time & 4 & (0, 197) (197, 264) (264, 915) (915, 15604) ms & Fig. \ref{fig:9_pennylane_optax_adamw_state_Sim14Ansatz_execution_time_20_20_001_4} \\
    \hline
    cardinality & 4 & (0, 7) (8, 33181) (33182, 635188) (635189, 102540525) rows & Fig. \ref{fig:1_pennylane_optax_adamw_state_Sim14Ansatz_cardinality_20_20_01_4} \\
    \hline
    cost & 4 & (0.0, 67132) (67139, 83735) (83953, 158348) (158590, 1425540) psql cost & Fig. \ref{fig:5_pennylane_optax_adamw_state_Sim14Ansatz_cost_20_20_01_4} \\
    \hline
    \hline
    execution time & 8 & (0, 156) (156, 202) (202, 226) (226, 256) (257, 406) (406, 904) (907, 2040) (2048, 18767) ms & Fig. \ref{fig:9_pennylane_optax_adamw_state_Sim14Ansatz_execution_time_20_20_01_8} \\
    \hline
    cardinality & 8 & (0, 0) (0, 0) (0, 4829) (5536, 29757) (33181, 143849) (150166, 577460) (635188, 1572370) (1626450, 102555254) rows & Fig. \ref{fig:1_pennylane_optax_adamw_state_Sim14Ansatz_cardinality_20_20_01_8} \\
    \hline
    cost & 8 & (0, 63565) (63770, 67139) (67158, 70690) (71128, 84086) (84276, 108703) (108703, 159742) (160257, 580270) (580470, 1425540)  psql cost & Fig. \ref{fig:5_pennylane_optax_adamw_state_Sim14Ansatz_cost_20_20_01_8} \\
    \hline
    \end{tabular}%
    }
    \caption{Statistics about the classification tasks. Classes are created so that every class contains approximately an equal number of queries.}
    \label{tab:statistics2}
\end{table*}

%% file: sections/evaluation.tex
\section{Evaluation}
All of the results are presented considering the new iterative training pipeline. We started the training with a small number of circuits and incrementally increased the number. The number of test and validation circuits is the same as the training circuits until we have included all the test and validation circuits ($\sim 130$ circuits). The number of these circuits is on the x-axis. The y-axis presents the accuracy.

\input{sections/binary_classification}
\input{sections/four_class_classification}
\input{sections/eight_class_classification}

\subsubsection*{Comparison to state-of-the-art databases.} To demonstrate the model's performance against state-of-the-art databases, we migrated the Internet Movie Database from PostgreSQL to MySQL and SQL Server. Since execution time and cost depend on hardware and databases, and our model is trained on PostgreSQL data, the comparison of execution times and costs is not possible. However, cardinalities remain consistent across databases, allowing comparison.

To perform the comparison, we interpret the database estimators as query classifiers. The selected databases implement cardinality estimators, which estimate cardinalities without executing the queries. We used the estimators to collect estimates for the training, validation, and test queries. Then, we used these estimates to classify the queries according to the same classes that we defined in Table~\ref{tab:statistics2}. Table \ref{tab:db_comparison_results} shows the results of this classification task.

\input{plots/results_from_databases}

The comparison's results are positive, considering that most quantum machine learning models have solved small prototypical use cases, which would be trivial for classical systems. Our model matches the accuracy of the classical systems in this use case. This is a positive indicator that this model might have beneficial features to perform well in real-life problems when the quantum systems scale up.

\subsubsection*{Limitations} One of the biggest challenges in this method is that the model is a classifier while the predicting metrics for queries is ideally a regression problem. Database optimizers provide a numerical estimate instead of a classification, which is essential for effective query processing. Theoretically, this problem can be solved by increasing the number of classes, which will ideally lead to better accuracy. The number of classes increases exponentially since every added target qubit doubles the number of classes. On the other hand, this requires more training data, larger models, and an increasing number of shots, which will likely lead to a more complex quantum machine learning setup to train.

The current evaluation focused on simple SQL queries having only a small number of join and filtering clauses, which map to around 20 qubits. Encoding the original queries in the join order benchmark requires around 70-100 qubits. As the number of qubits increases, training the model becomes more challenging.

Finally, post-selection is an expensive operation. For example, learned measurements \cite{GarciaPerez_Rossi_Sokolov_Tacchino_Barkoutsos_Mazzola_Tavernelli_Maniscalco_2021} could provide an alternative layer at the end of the circuits. Also, more sophisticated mid-circuit measurement-based schemes or additive approximation of an amplitude encoding could solve the expense of post-selection \cite{Arad_Landau_2010}.

%% file: sections/binary_classification.tex
\subsubsection*{Binary classification results}
\begin{figure*}
    \begin{subfigure}[t]{0.32\textwidth}
    \centering
    \includegraphics[width=\textwidth]{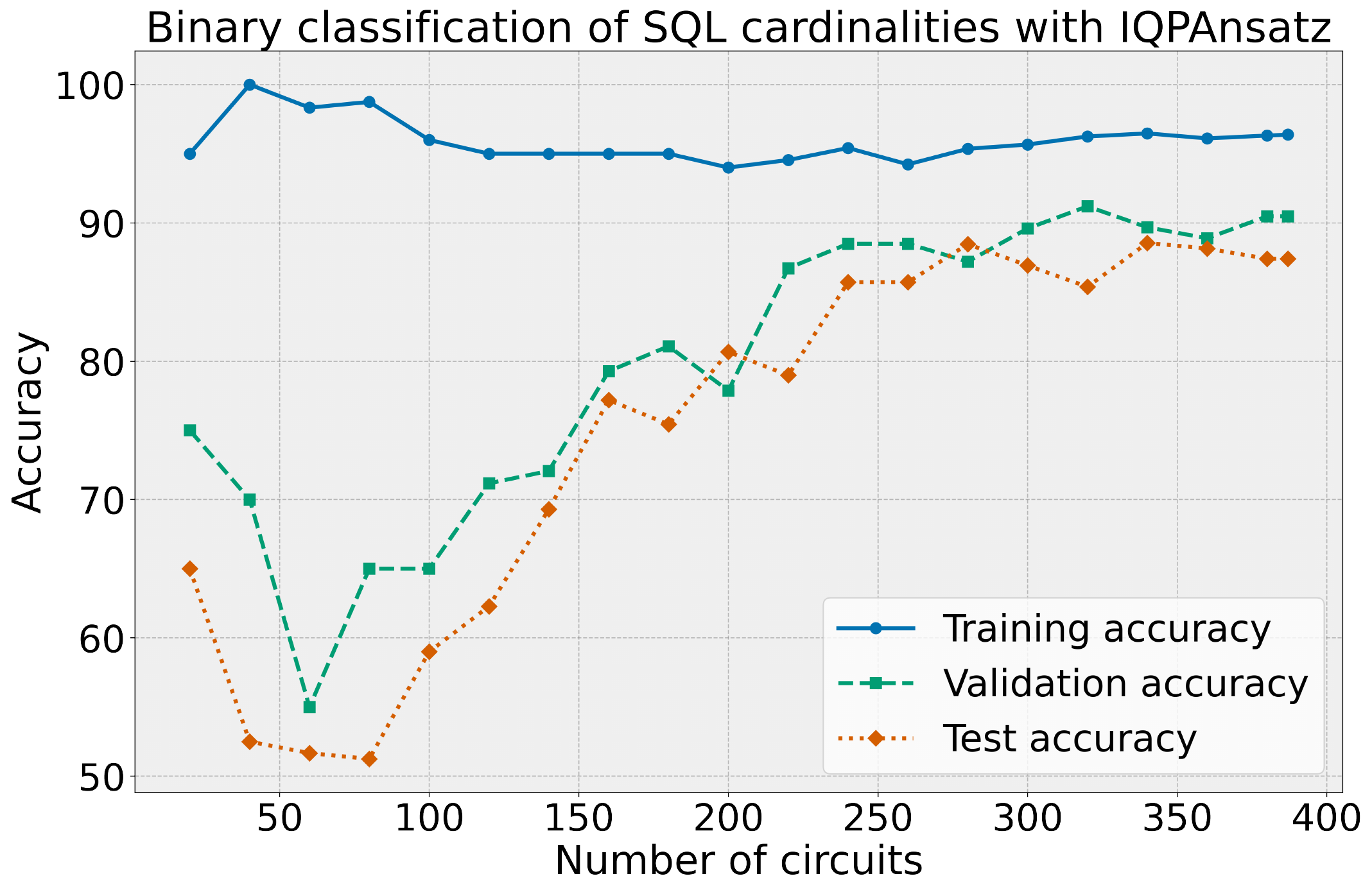}
    \caption{Binary classification for cardinalities with IQPAnsatz}
    \label{fig:1_pennylane_optax_adamw_state_IQPAnsatz_cardinality_20_20_001_2}
  \end{subfigure}
  \begin{subfigure}[t]{0.32\textwidth}
    \centering
    \includegraphics[width=\textwidth]{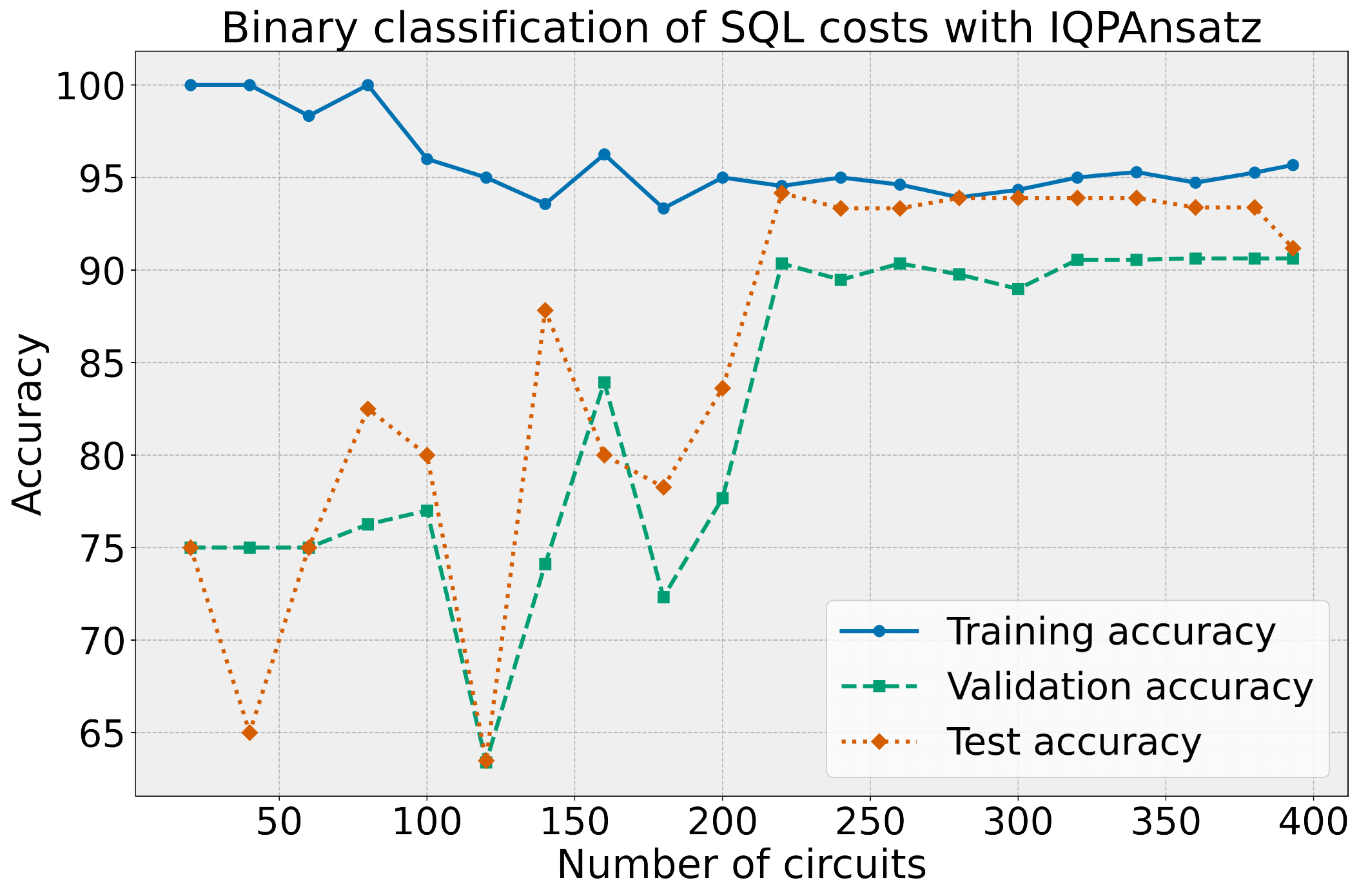}
    \caption{Binary classification for costs with IQPAnsatz}
    \label{fig:5_pennylane_optax_adamw_state_IQPAnsatz_cost_20_20_001_2}
  \end{subfigure}
  \begin{subfigure}[t]{0.32\textwidth}
    \centering
    \includegraphics[width=\textwidth]{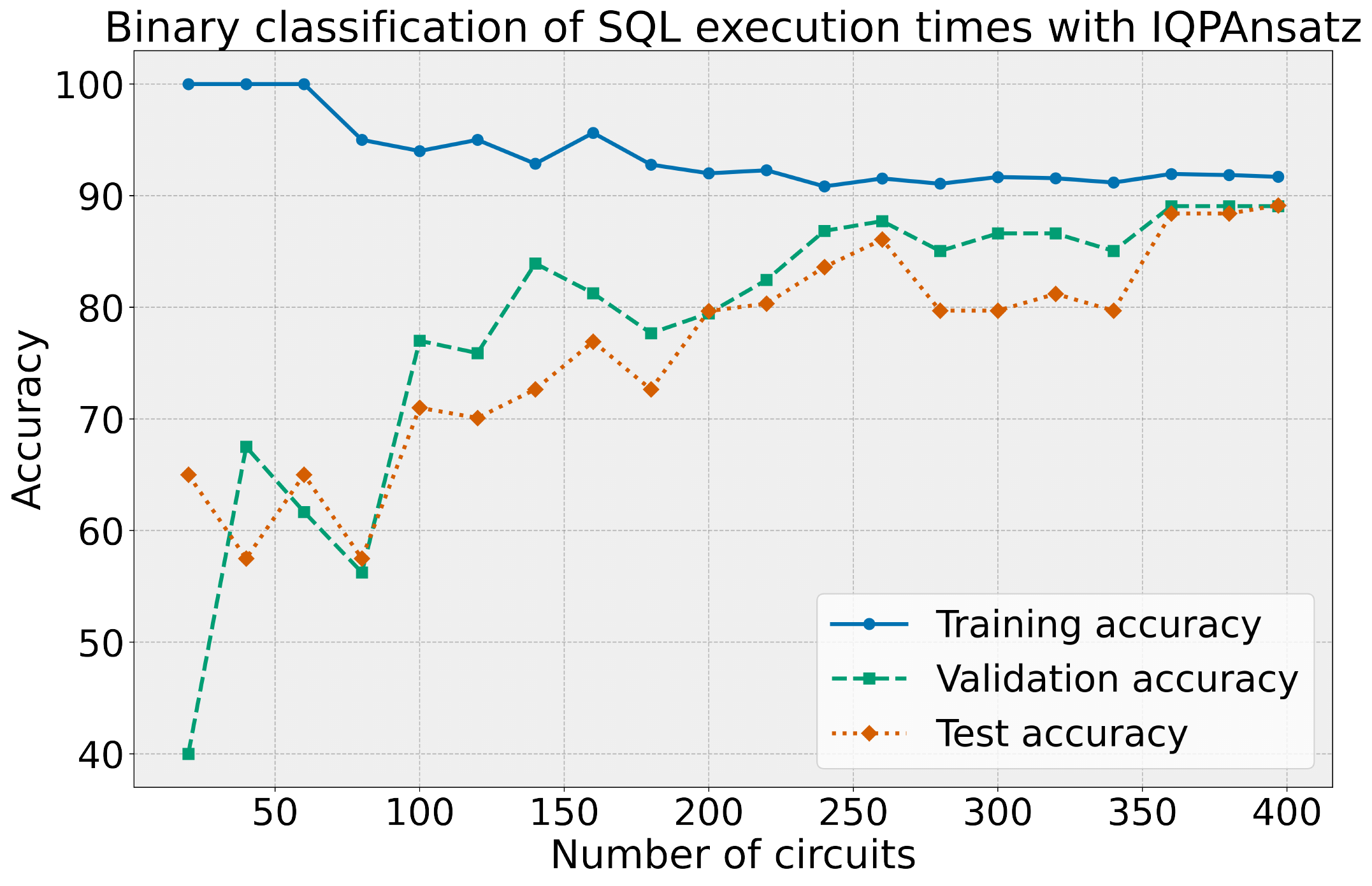}
    \caption{Binary classification for execution times with IQPAnsatz}
    \label{fig:9_pennylane_optax_adamw_state_IQPAnsatz_execution_time_20_20_001_2}
  \end{subfigure}
\end{figure*}

Quantum machine learning is still in its early stages, where even binary classification remains a central and significant challenge \cite{bowles2024betterclassicalsubtleart}. The results from the binary classification task in all three cases are excellent. The accuracy of the cardinality estimations are in Fig.~\ref{fig:1_pennylane_optax_adamw_state_IQPAnsatz_cardinality_20_20_001_2}, costs in Fig.~\ref{fig:5_pennylane_optax_adamw_state_IQPAnsatz_cost_20_20_001_2} and execution times in Fig.~\ref{fig:9_pennylane_optax_adamw_state_IQPAnsatz_execution_time_20_20_001_2}. Our binary classification accuracy is competitive against the similar model in QNLP \cite{Lorenz_Pearson_Meichanetzidis_Kartsaklis_Coecke_2021}, meaning that this QNLP-based model generalizes outside of natural language processing. 

\begin{figure*}
    \begin{subfigure}[t]{0.32\textwidth}
    \centering
    \includegraphics[width=\textwidth]{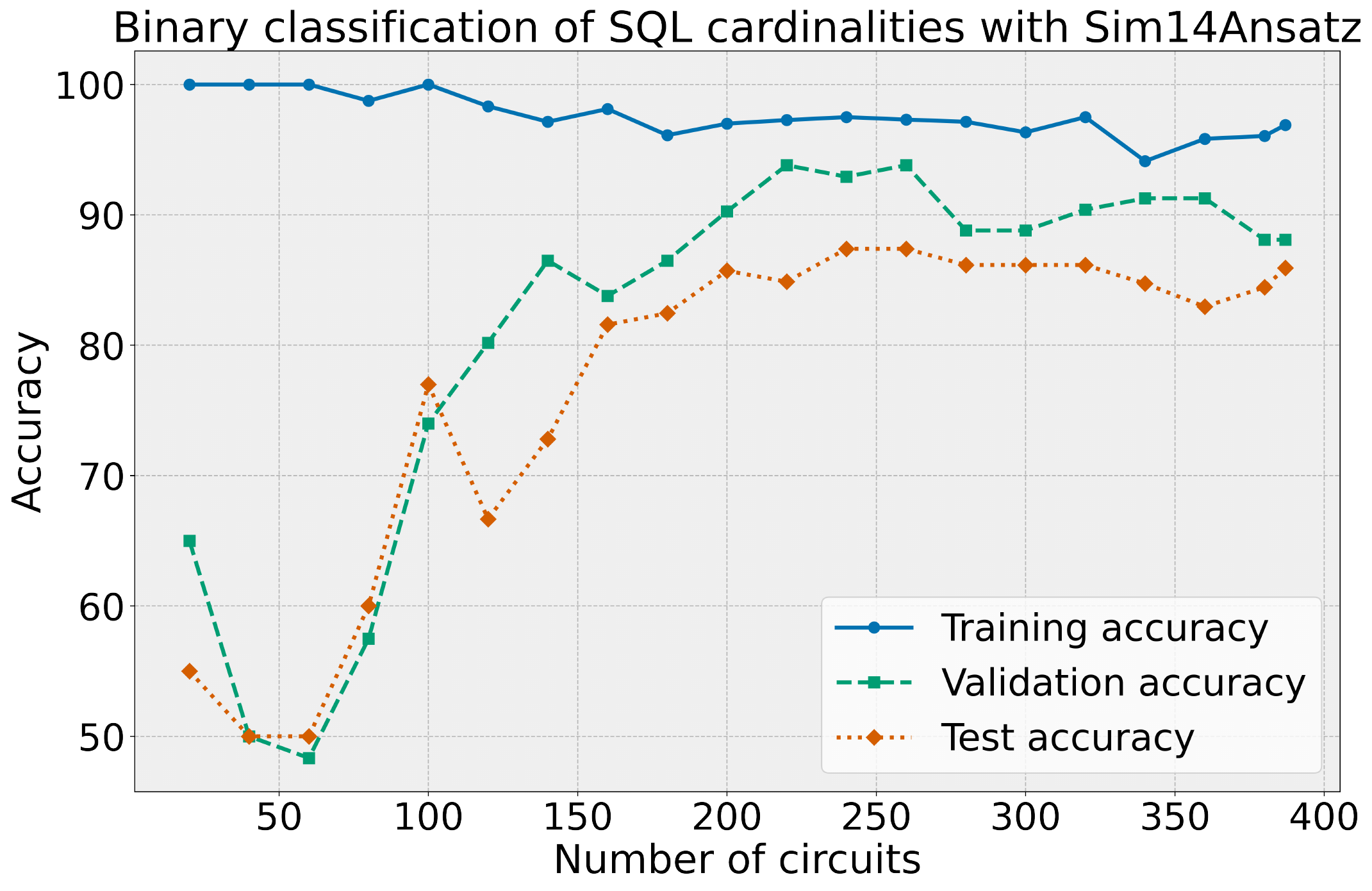}
    \caption{Binary classification for cardinalities with Sim14Ansatz}
    \label{fig:1_pennylane_optax_adamw_state_Sim14Ansatz_cardinality_20_20_001_2}
  \end{subfigure}
  \begin{subfigure}[t]{0.32\textwidth}
    \centering
    \includegraphics[width=\textwidth]{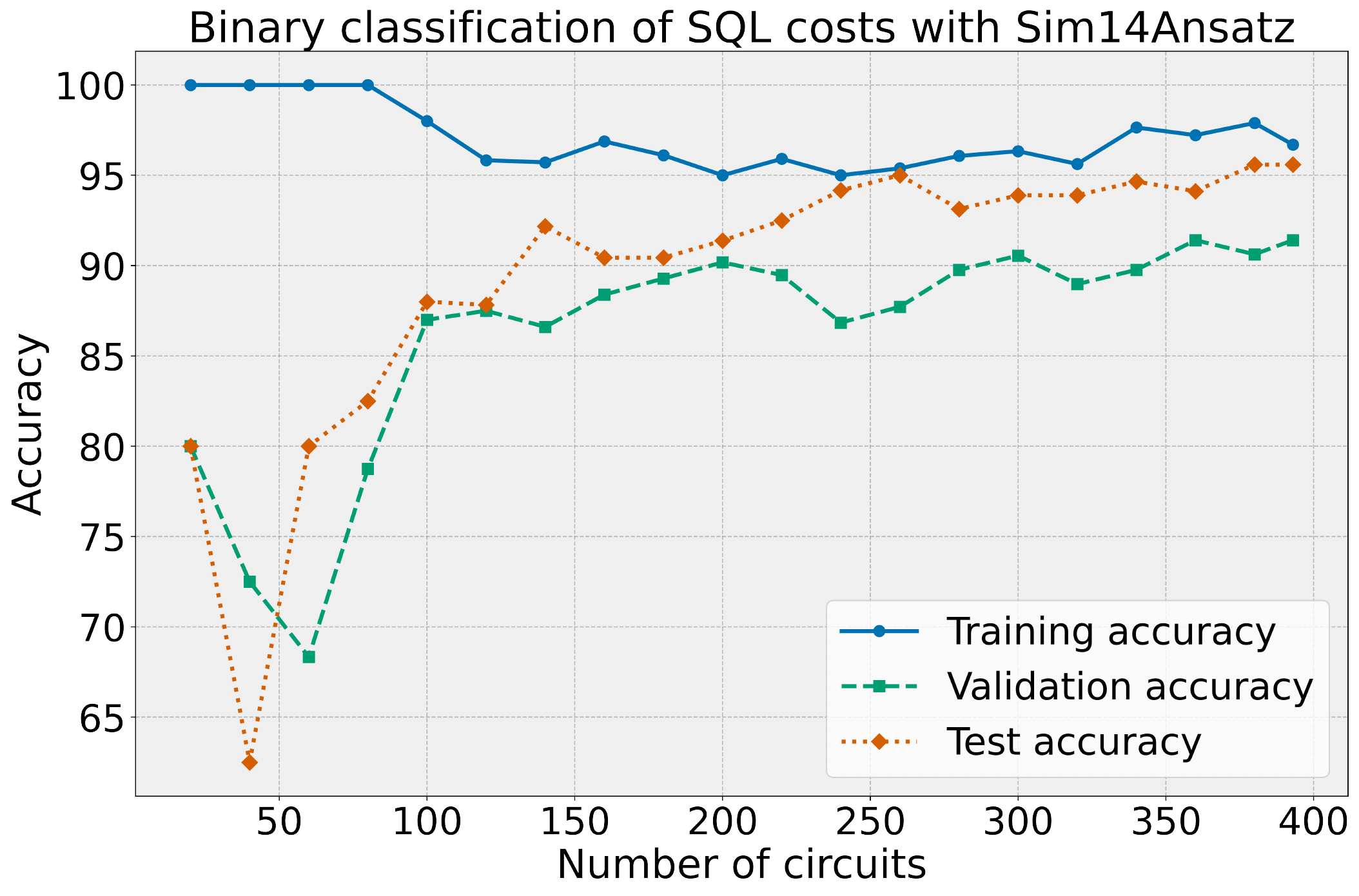}
    \caption{Binary classification for costs with Sim14Ansatz}
    \label{fig:5_pennylane_optax_adamw_state_Sim14Ansatz_cost_20_20_001_2}
  \end{subfigure}
  \begin{subfigure}[t]{0.32\textwidth}
    \centering
    \includegraphics[width=\textwidth]{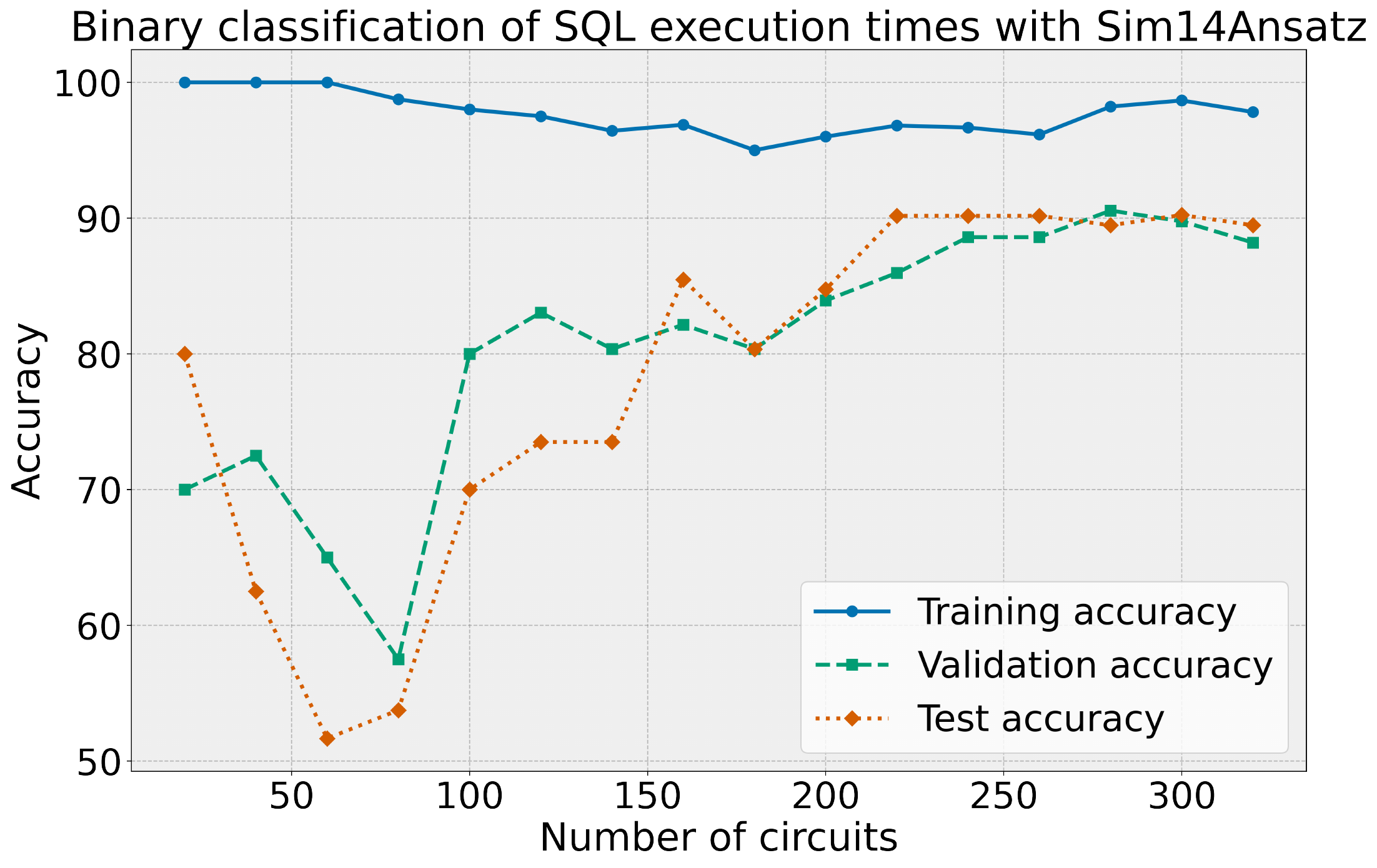}
    \caption{Binary classification for execution times with Sim14Ansatz}
    \label{fig:9_pennylane_optax_adamw_state_Sim14Ansatz_execution_time_20_20_001_2}
  \end{subfigure}
\end{figure*}

We also studied the performance of two different circuit layouts: IQPAnsatz and Sim14Ansatz. Sim14Ansatz refers to the circuit template number 14 in \cite{Sim_Johnson_Aspuru_Guzik_2019}. Considering the corresponding results using Sim14Ansatz in Fig.~\ref{fig:1_pennylane_optax_adamw_state_Sim14Ansatz_cardinality_20_20_001_2}, Fig.~\ref{fig:5_pennylane_optax_adamw_state_Sim14Ansatz_cost_20_20_001_2} and Fig.~\ref{fig:9_pennylane_optax_adamw_state_Sim14Ansatz_execution_time_20_20_001_2}, we can see that Sim14Ansatz might reach slightly higher accuracy compared to IQPAnsatz. This is expected, as Sim14Ansatz is more expressible. However, its higher expressibility also makes it harder to train, which suggests that the simpler IQPAnsatz may be sufficient for binary classification tasks.

%% file: sections/four_class_classification.tex
\subsubsection*{4-class classification results}
\begin{figure*}
  \begin{subfigure}[t]{0.32\textwidth}
    \centering
    \includegraphics[width=\textwidth]{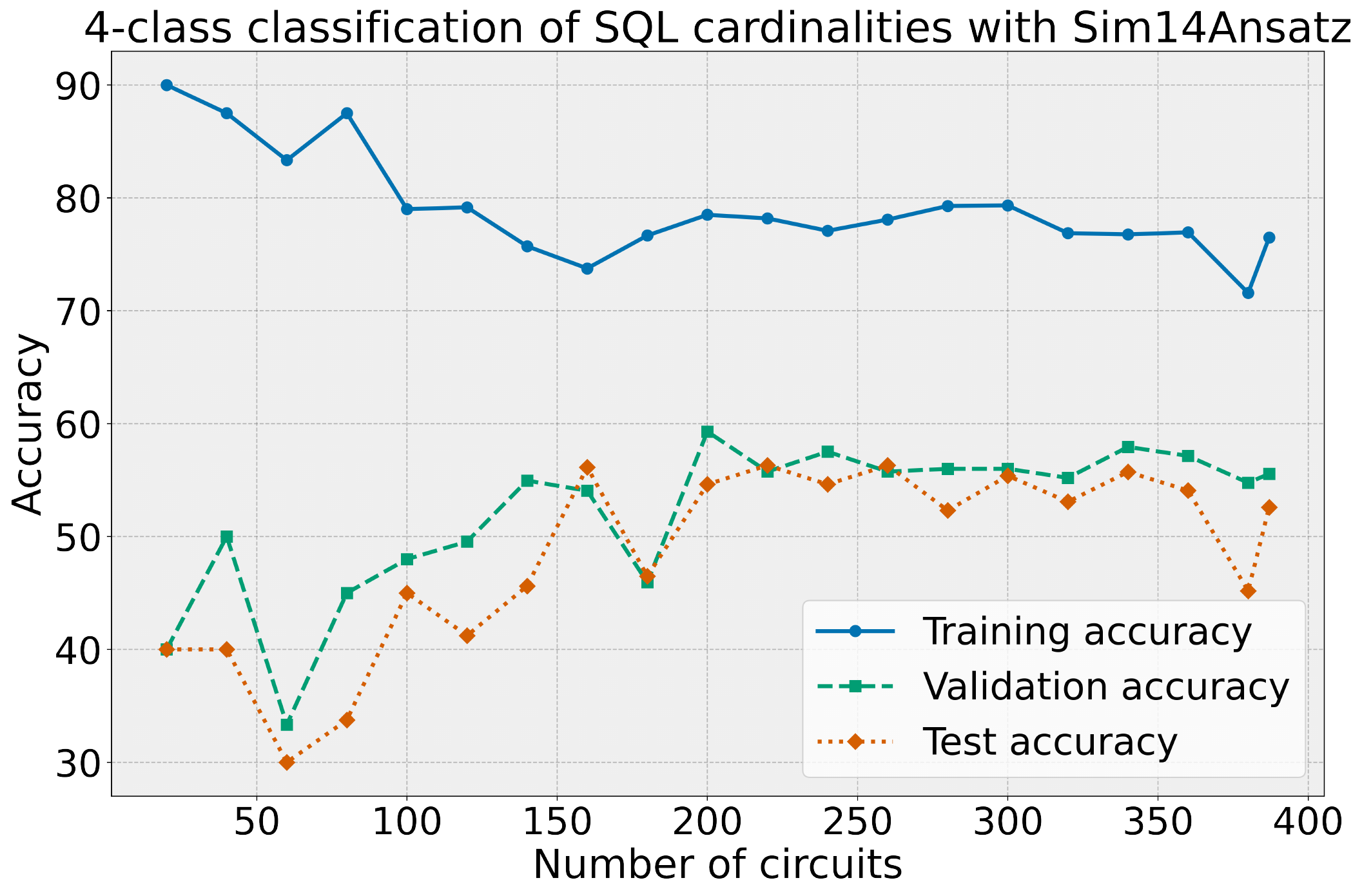}
    \caption{4-class classification for cardinalities with Sim14Ansatz}
    \label{fig:1_pennylane_optax_adamw_state_Sim14Ansatz_cardinality_20_20_01_4}
  \end{subfigure} \hspace{0.5em}
  \begin{subfigure}[t]{0.32\textwidth}
    \centering
    \includegraphics[width=\textwidth]{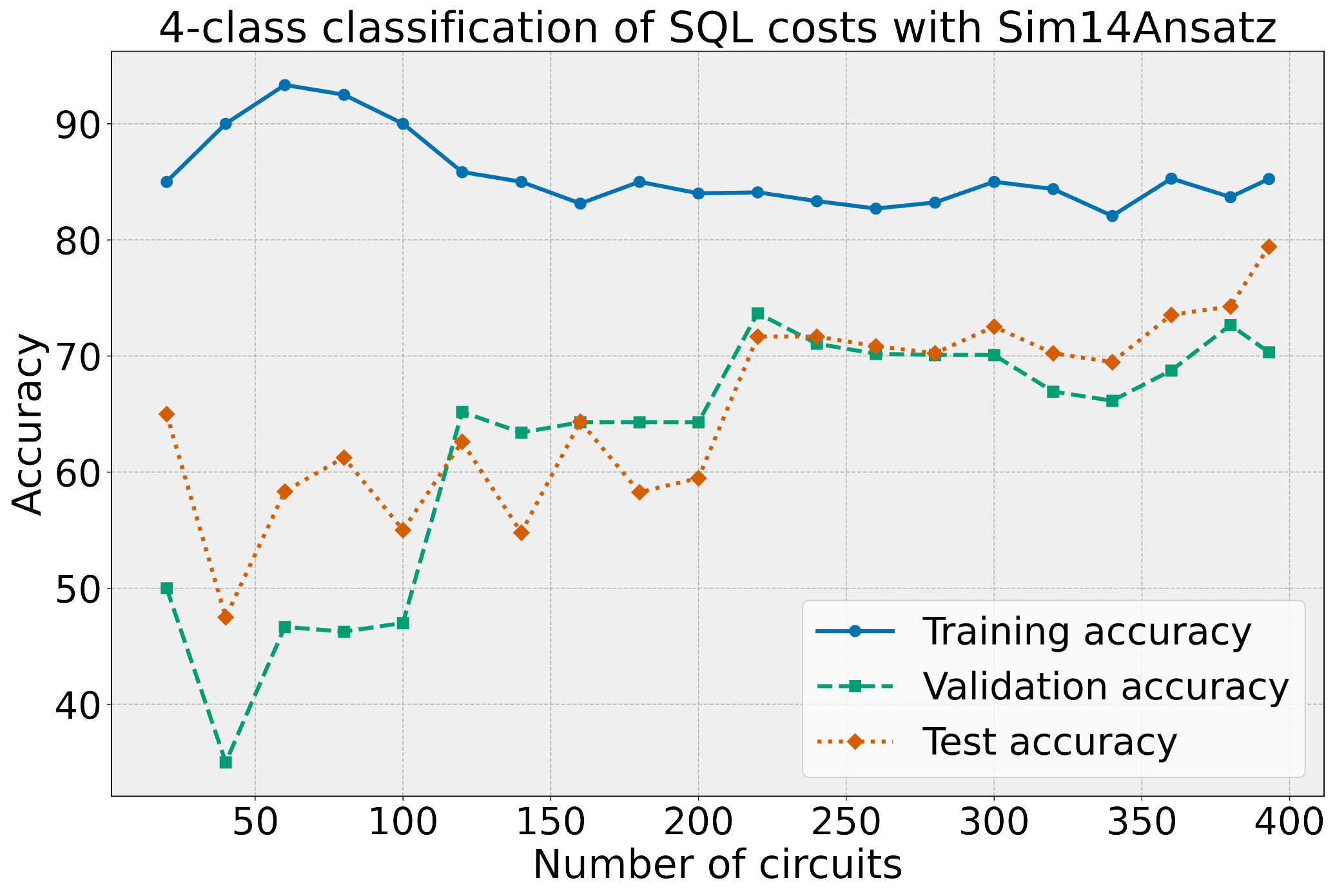}
    \caption{4-class classification for costs with Sim14Ansatz}
    \label{fig:5_pennylane_optax_adamw_state_Sim14Ansatz_cost_20_20_01_4}
  \end{subfigure} \hspace{0.5em}
  \begin{subfigure}[t]{0.32\textwidth}
    \centering
    \includegraphics[width=\textwidth]{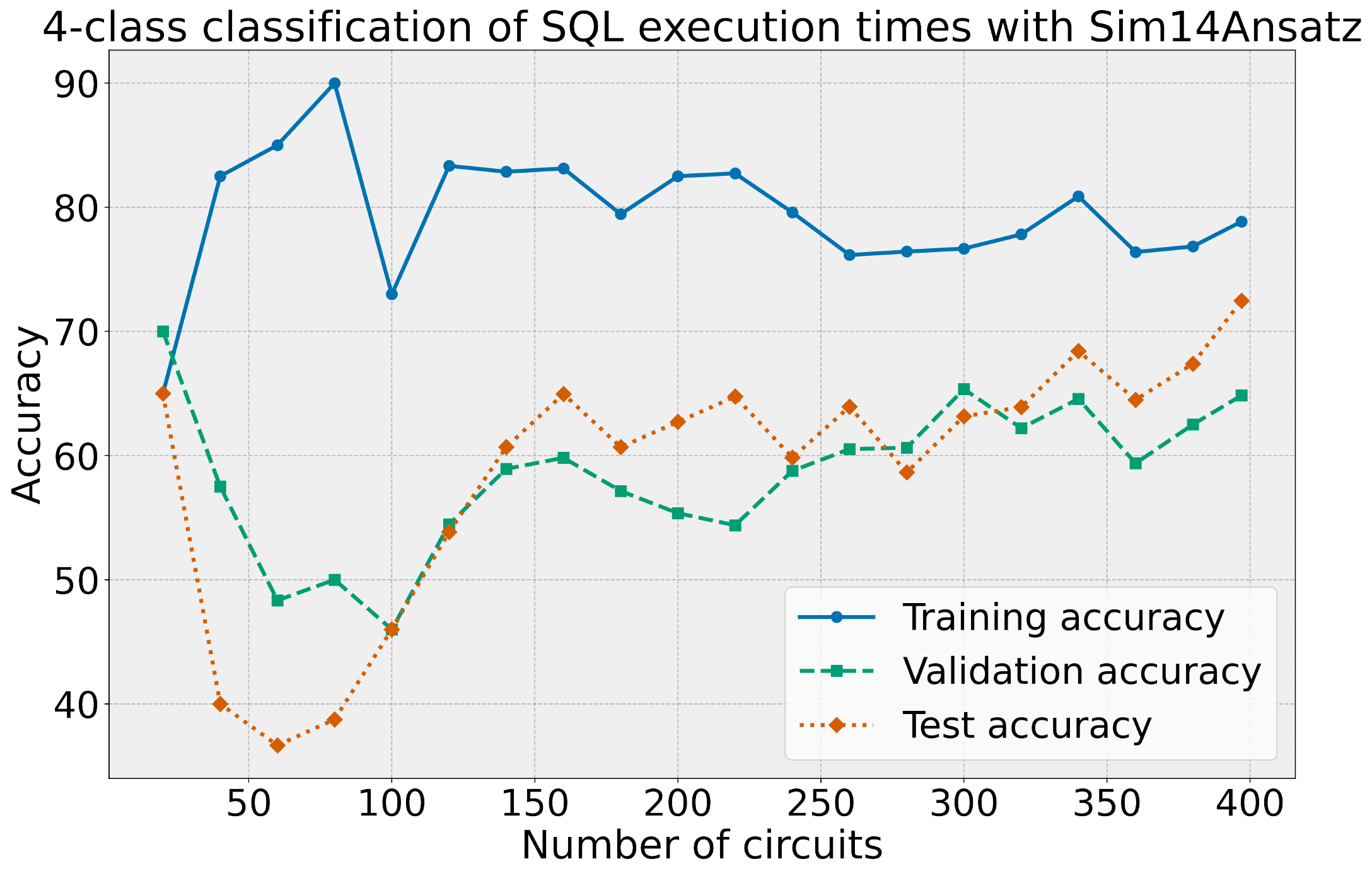}
    \caption{4-class classification for execution times with Sim14Ansatz}
    \label{fig:9_pennylane_optax_adamw_state_Sim14Ansatz_execution_time_20_20_001_4}
  \end{subfigure}
\end{figure*}

Since quantum machine learning might not often scale beyond binary classification, we have obtained good and usable results from the 4-class classification task. The previous QNLP research did not consider 4-class classification using this model. Fig.~\ref{fig:1_pennylane_optax_adamw_state_Sim14Ansatz_cardinality_20_20_01_4} shows the results from cardinality estimations, Fig.~\ref{fig:5_pennylane_optax_adamw_state_Sim14Ansatz_cost_20_20_01_4} from cost estimations and Fig.~\ref{fig:9_pennylane_optax_adamw_state_Sim14Ansatz_execution_time_20_20_001_4} from execution time estimations.

\begin{figure}
    \centering
    \includegraphics[width=0.65\linewidth]{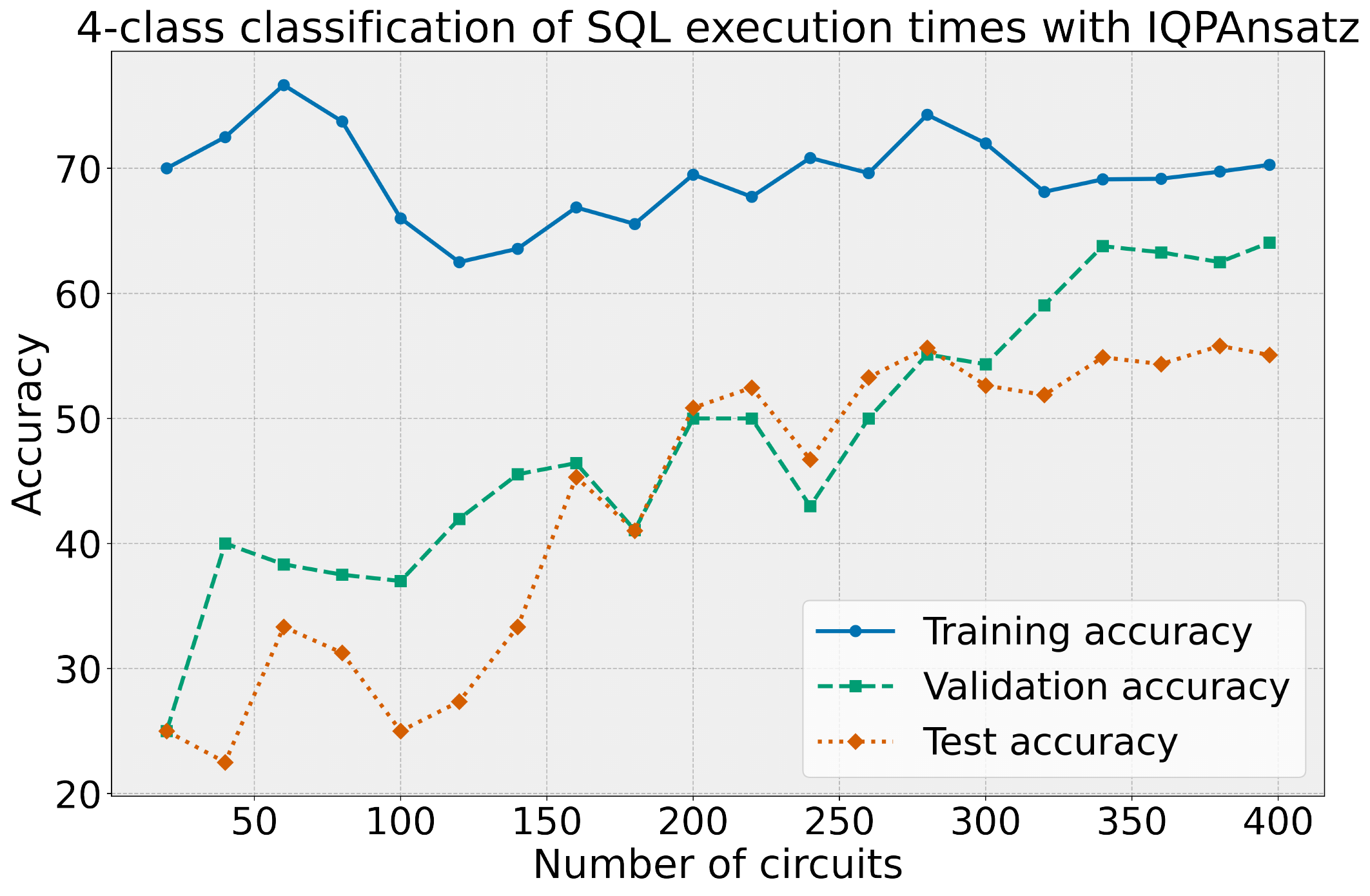}
    \caption{Comparing IQPAnsatz to Sim14Ansatz reveals the performance difference between these circuit layouts.}
    \label{fig:9_pennylane_optax_adamw_state_IQPAnsatz_execution_time_20_20_001_4}
\end{figure}

Fig.~\ref{fig:9_pennylane_optax_adamw_state_IQPAnsatz_execution_time_20_20_001_4} demonstrates the performance of the IQPAnsatz in the same 4-class classification tasks, which does not match the performance of the Sim14Ansatz. We have excluded this comparison for the 8-class classification, as the model showed minimal learning when using the IQPAnsatz. Thus, 8-class results should use the Sim14Ansatz.

%% file: sections/eight_class_classification.tex
\subsubsection*{8-class classification results}

\begin{figure*}
  \begin{subfigure}[t]{0.32\textwidth}
    \centering
    \includegraphics[width=\textwidth]{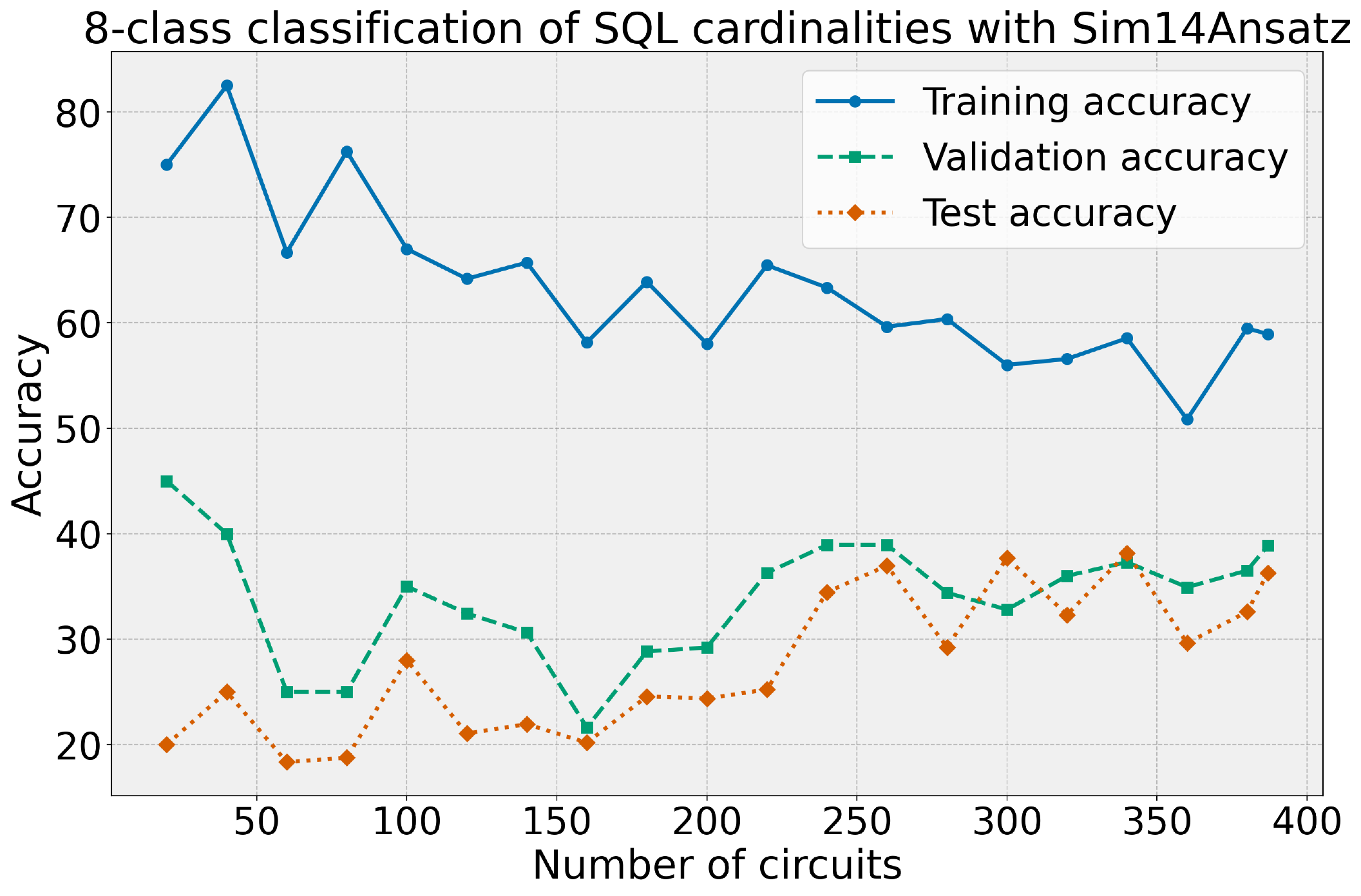}
    \caption{8-class classification for cardinalities with Sim14Ansatz}
    \label{fig:1_pennylane_optax_adamw_state_Sim14Ansatz_cardinality_20_20_01_8}
  \end{subfigure}
  \begin{subfigure}[t]{0.32\textwidth}
    \centering
    \includegraphics[width=\textwidth]{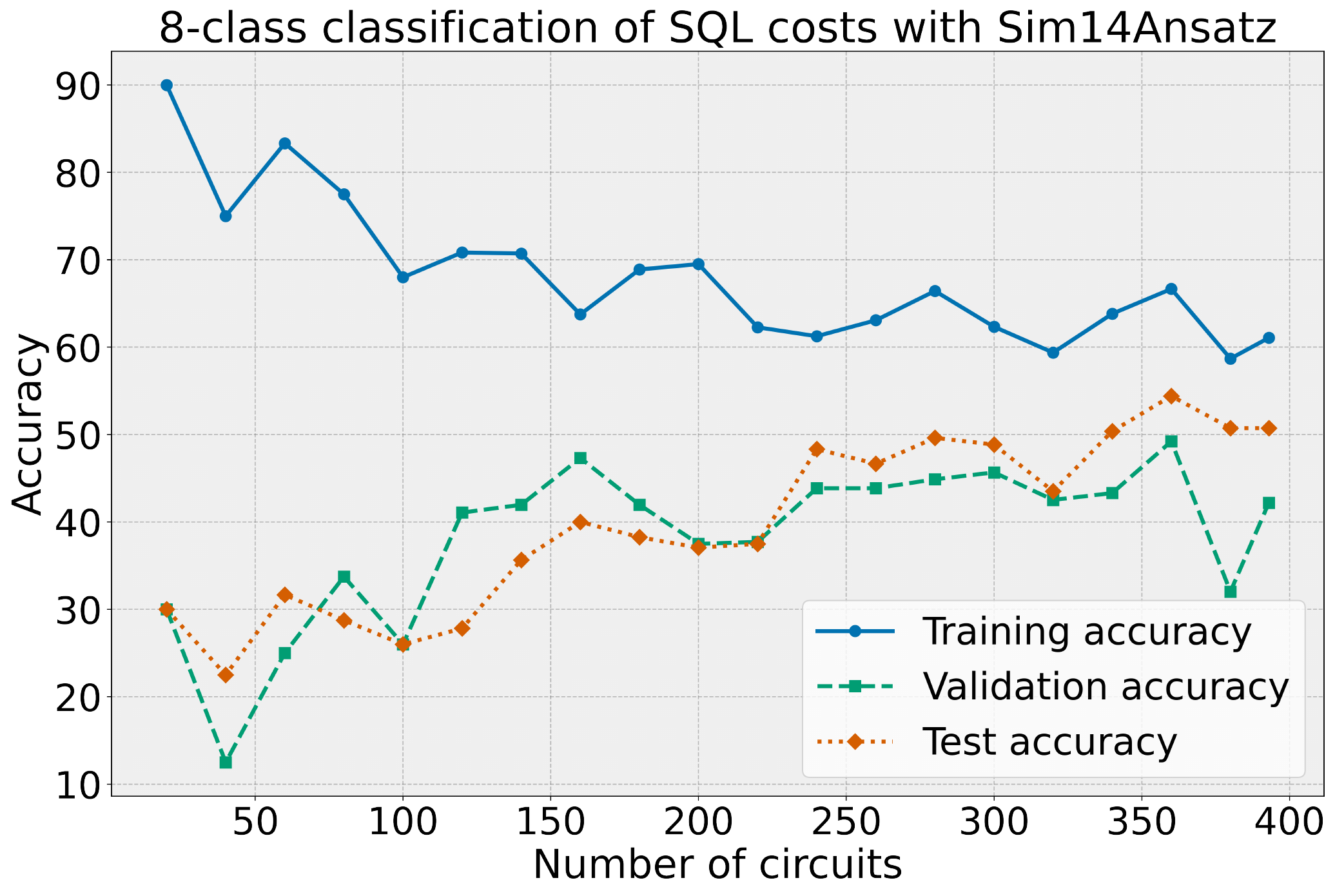}
    \caption{8-class classification for costs with Sim14Ansatz}
    \label{fig:5_pennylane_optax_adamw_state_Sim14Ansatz_cost_20_20_01_8}
  \end{subfigure}
  \begin{subfigure}[t]{0.32\textwidth}
    \centering
    \includegraphics[width=\textwidth]{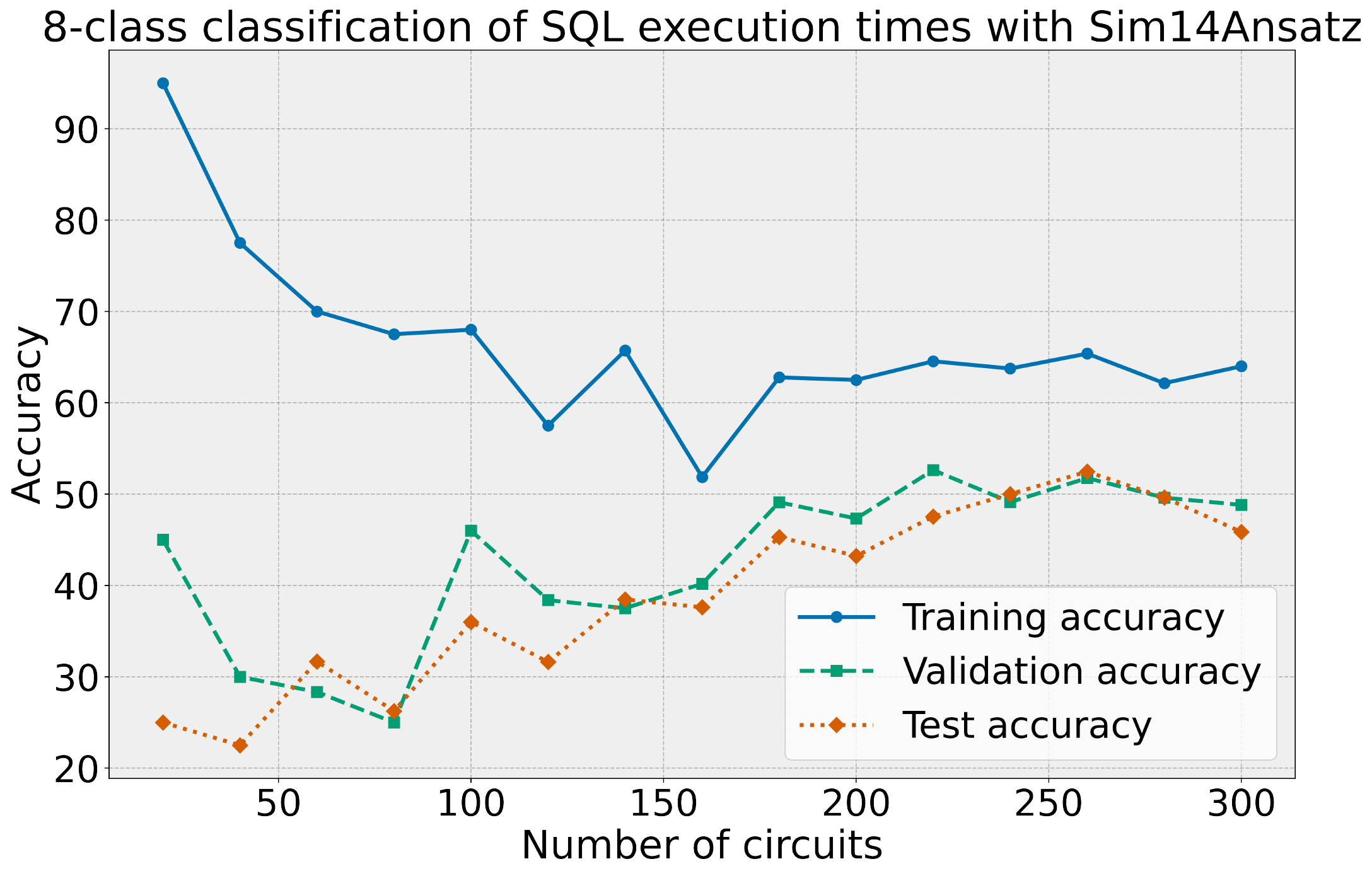}
    \caption{8-class classification for execution times with Sim14Ansatz}
    \label{fig:9_pennylane_optax_adamw_state_Sim14Ansatz_execution_time_20_20_01_8}
  \end{subfigure}
\end{figure*}

As the results in Fig.~\ref{fig:5_pennylane_optax_adamw_state_Sim14Ansatz_cost_20_20_01_8}, Fig.~\ref{fig:1_pennylane_optax_adamw_state_Sim14Ansatz_cardinality_20_20_01_8}, and Fig.~\ref{fig:9_pennylane_optax_adamw_state_Sim14Ansatz_execution_time_20_20_01_8} show, the performance of the model in 8-class classification is satisfactory, but it is possible that even the Sim14Ansatz is not expressive enough for this task. On the other hand, by making the circuits more expressive, the training becomes more challenging. At the same time, the size of the circuits reached the limit where they are simulable, making the model hard to train without extensive classical resources \cite{bowles2024betterclassicalsubtleart}. Unfortunately, even if we had perfect quantum computers, quantum machine learning would suffer from problems that prevent training large models \cite{Ragone_Bakalov_Sauvage_Kemper_Ortiz_Marrero_Larocca_Cerezo_2024}. Overcoming these problems requires novel solutions from future quantum machine learning techniques, and we most likely need to develop methods to evaluate gradients on quantum computers.

%% file: plots/results_from_databases.tex
\begin{table*}[t]
        \centering
        \resizebox{0.8\textwidth}{!}{%
    \begin{tabular}{|c|c|c|c|c|c|}
    \hline
    Database & \#classes & Accuracy on training queries & Accuracy on validation queries & Accuracy on test queries & Average accuracy \\
    \hline
    PostgreSQL & 2 & 81.65 & 83.46 & 83.94 & 83.02 \\
    \hline
    MySQL & 2 & 47.29 & 49.62 & 50.36 & 49.09 \\
    \hline
    SQL Server & 2 & 90.96 & 92.48 & 91.97 & 91.80 \\
    \hline
    SQL2Circuits & 2 & 96.38 & 90.47 & 87.40 & 91.42 \\
    \hline
    \hline
    PostgreSQL & 4 & 57.62 & 67.67 & 62.04 & 62.45 \\
    \hline
    MySQL & 4 & 27.65 & 30.83 & 33.58 & 30.68 \\
    \hline
    SQL Server & 4 & 61.76 & 68.42 & 61.31 & 63.83 \\
    \hline
    SQL2Circuits & 4 & 76.48 & 55.55 & 52.59 & 61.54 \\
    \hline
    \hline
    PostgreSQL & 8 & 44.17 & 45.99 & 55.64 & 48.60 \\
    \hline
    MySQL & 8 & 12.92 & 14.29 & 13.87 & 13.69 \\
    \hline
    SQL Server & 8 & 47.80 & 55.64 & 45.26 & 49.57 \\
    \hline
    SQL2Circuits & 8 & 58.91 & 38.89 & 36.29 & 44.70 \\
    \hline
    \end{tabular}%
    }
    \caption{The accuracy of the SQL2Circuits model compared to state-of-the-art database cardinality estimators when both are employed as query classifiers, using the same database and queries.}
    \label{tab:db_comparison_results}
\end{table*}

%% file: sections/analysis_of_model.tex
\section{Analysis of model}
We perform a theoretical analysis of the developed quantum machine learning model, whose main findings are presented in Fig.~\ref{fig:analysis_of_model}. We are unaware that previous research would have analyzed this model considering these two quantum mechanical metrics. Computing these metrics is essential to understand how technically feasible it is to train and execute this model on real hardware.

\input{results/expressibility_main}

\subsubsection*{Expressibility}
Intuitively, in the context of a single circuit with a single qubit, expressibility refers to the property of how well the circuit can express different pure quantum states, i.e., how well it covers the surface of the Bloch sphere \cite{Sim_Johnson_Aspuru_Guzik_2019}. When executed with randomly initialized parameters, the circuit's distribution is compared to the Haar distribution, which is the uniform distribution over the state space. To compare these distributions, we compute the Kullback-Leibler divergence (also known as relative entropy) \cite{Kullback_Leibler_1951}, a measure commonly used in classical machine learning.
\begin{displaymath}
    \mathrm{Exp} := D_{KL}(P_{PQC}(F;S)\ ||\ P_{\mathrm{Haar}}(F)),
\end{displaymath}
where $P_{PQC}(F;S)$ is the estimated probability distribution of fidelities when randomly sampling states from the circuits in the model, and $D_{KL}$ is the Kullback-Leibler divergence. The set $S$ refers to the uniformly randomly sampled parameter configurations.


Expressibility for the training circuits for the binary classification of execution times is visualized in Figure \ref{fig:expressibility_1}. The circuit architectures for cardinality and cost estimations are similar, so the calculations can be generalized to these cases. The expressibility of the model is good because its Kullback-Leibler divergence value is 0.017, which is close to those divergence values that correspond to the well-performing circuits ($< 0.02$) in the previous evaluation \cite{Sim_Johnson_Aspuru_Guzik_2019}. 

Furthermore, we studied the expressibility of the circuits used in 4-class and 8-class classification. The histograms and the corresponding Haar distribution are in Figures \ref{fig:expressibility_2} and \ref{fig:expressibility_3}. The fact that the shape of the histograms follows the Haar distribution means that the distributions are close to each other. The expressibility of the circuits decreases, being $0.032$ and $0.056$, respectively. These values are not yet considered poor, but they approach the values for which circuits were found to be sub-optimal ($\mathrm{Exp} > 0.09$) \cite{Sim_Johnson_Aspuru_Guzik_2019}. This suggests that we may need to modify the circuit layout to represent the classification problem effectively when the number of classes increases. This also aligns with the results we obtained from evaluating the model: its performance started to decrease when the number of classes grew, as Figures \ref{fig:5_pennylane_optax_adamw_state_Sim14Ansatz_cost_20_20_01_8}, \ref{fig:1_pennylane_optax_adamw_state_Sim14Ansatz_cardinality_20_20_01_8}, and \ref{fig:9_pennylane_optax_adamw_state_Sim14Ansatz_execution_time_20_20_01_8} demonstrate.

\subsubsection*{Entangling capability}
The entangling capability of a circuit is related to its property of efficiently representing solutions utilizing entangled states, preferably with a small number of gates. This work calculates the entangling capability with the Meyer-Wallach entangling measure \cite{Sim_Johnson_Aspuru_Guzik_2019}. Computing the measure is relatively technical and can be found in \cite{Brennen_2003}. The key concept is that two qubits can exist in two extremes: completely separate (represented by $0$) or maximally entangled (represented by $1$). An example of a maximally entangled state is the Bell state. This principle extends to systems with multiple qubits.

The entangling capability for the training circuits in binary classification of execution times is visualized in Figure \ref{fig:entangling_capability_1}. The average Meyer-Wallach entangling measure over the training circuits for binary classification of execution times is $0.517$. Compared to the entangling capabilities reported in \cite{Sim_Johnson_Aspuru_Guzik_2019}, the circuits in our model have entangling capabilities equivalent to those of circuits with just a single entangling layer. The entangling capability of the circuits in our model is on the scale ($0.4-0.7$) of the favorable circuits identified in \cite{Sim_Johnson_Aspuru_Guzik_2019}. 

After calculating the average entangling measure for the circuits representing 4-class and 8-class classification tasks, we did not observe a significant difference in the entanglement results between the binary classification circuits. The values were $0.554$ and $0.570$, respectively. The histograms are similar to Figure \ref{fig:entangling_capability_1}, so we have omitted them. These results are expected because the circuit layout remains consistent across binary, 4-class, and 8-class classification tasks. A similar circuit layout creates similar entanglement between the qubits, making the entanglement measurements close.

%% file: results/expressibility_main.tex
\begin{figure*}
  \begin{subfigure}[t]{0.24\textwidth}
    \centering
    \begin{adjustbox}{width=\textwidth}
    \input{results/expresibility_1}
    \end{adjustbox}
    \caption{The KL divergence is 0.017 for the binary classification}
    \label{fig:expressibility_1}
  \end{subfigure}
  \begin{subfigure}[t]{0.24\textwidth}
    \centering
    \begin{adjustbox}{width=\textwidth}
    \input{results/expressibility_2}
    \end{adjustbox}
    \caption{The KL divergence is 0.032 for the 4-class classification}
    \label{fig:expressibility_2}
  \end{subfigure}
  \begin{subfigure}[t]{0.24\textwidth}
    \centering
    \begin{adjustbox}{width=\textwidth}
    \input{results/expressibility_3}
    \end{adjustbox}
    \caption{The KL divergence is 0.056 for the 8-class classification}
    \label{fig:expressibility_3}
  \end{subfigure}
    \begin{subfigure}[t]{0.24\textwidth}
    \centering
    \begin{adjustbox}{width=\textwidth}
    \input{results/engtangling_capability}
    \end{adjustbox}
    \caption{The average entangling capability is 0.517.}
    \label{fig:entangling_capability_1}
  \end{subfigure}
  \caption{Expressibility and entangling capability analysis of the model}
  \label{fig:analysis_of_model}
\end{figure*}

%% file: results/expresibility_1.tex
  \begin{tikzpicture}
  \begin{axis}[
      xlabel=Fidelity,
      ylabel=Probability,
      xtick={0.0, 0.25,0.5, 0.75, 1.0},
      ytick = {0.01, 0.0133, 0.015, 0.02},
      ybar,
      bar width=2pt,
      xmin=-0.1,
      xmax=1.1,
      ymajorgrids=true,
      grid style=dashed,
      tick label style={/pgf/number format/fixed}
      ]
    \addplot table[x=bin,y=fidelity, col sep=comma] {results/data/fidelity_execution_time_1_1_3_1_main.csv};
    \addplot[only marks, mark = -, mark size=2pt, red, samples=1000] {0.0133333333333};
     \end{axis}
\end{tikzpicture}

%% file: results/expressibility_2.tex
\begin{tikzpicture}
  \begin{axis}[
      xlabel=Fidelity,
      ylabel=Probability,
      xtick={0.0, 0.25,0.5, 0.75, 1.0},
      ytick = {0.001, 0.01, 0.02, 0.03, 0.04, 0.05, 0.06},
      xmin=-0.1,
      xmax=1.1,
      ymajorgrids=true,
      grid style=dashed,
      legend cell align={left},
      tick label style={/pgf/number format/fixed}
      ]
    \addplot[ybar, bar width=2pt, draw=blue] table[x=bin,y=fidelity, col sep=comma] {results/data/expressibility_2.csv};
    \addlegendentry{Expressibility historam}
    
    \addplot[
    color = red,
    mark = *,
    mark size=1pt
    ] table [x=bins_x, y=P_harr_hist, col sep=comma] {results/data/expressibility_2.csv};
    \addlegendentry{Haar distribution}
     \end{axis}
\end{tikzpicture}

%% file: results/expressibility_3.tex
\begin{tikzpicture}
  \begin{axis}[
      xlabel=Fidelity,
      ylabel=Probability,
      xtick={0.0, 0.25,0.5, 0.75, 1.0},
      ytick = {0.001, 0.025, 0.05, 0.075, 0.1},
      xmin=-0.1,
      xmax=1.1,
      ymajorgrids=true,
      grid style=dashed,
      legend cell align={left},
      tick label style={/pgf/number format/fixed}
      ]
    \addplot[ybar, bar width=2pt, draw=blue] table[x=bin,y=fidelity, col sep=comma] {results/data/expressibility_3.csv};
    \addlegendentry{Expressibility historam}
    
    \addplot[
    color = red,
    mark = *,
    mark size=1pt
    ] table [x=bins_x, y=P_harr_hist, col sep=comma] {results/data/expressibility_3.csv};
    \addlegendentry{Haar distribution}
     \end{axis}
\end{tikzpicture}

%% file: results/engtangling_capability.tex
  \begin{tikzpicture}
  \begin{axis}[
    xlabel = Circuit id,
    ylabel = Engtangling capability Q,
    xtick = {1, 100, 200, 300, 400, 500, 600, 670},
    ytick = {0, 0.25, 0.50, 0.75, 1},
    ybar,
    bar width=1pt,
    ymax = 1,
    ymin=0
  ]
    \addplot table[x=id,y=engtangling_capability, col sep=comma, fill] {results/data/mw_engtanglement_execution_time_1_1_3_1_main.csv};
     \end{axis}
\end{tikzpicture}

%% file: sections/conclusion_future_work.tex
\section{Conclusion and future work}

This work tackled the database problem of estimating cardinalities, costs, and execution times for SQL queries. We developed a quantum machine learning model based on quantum natural language processing \cite{Lorenz_Pearson_Meichanetzidis_Kartsaklis_Coecke_2021}. We implemented a robust category-theoretical encoding mechanism to translate SQL queries into parametrized circuits. The results on the binary classification task align with prior research in QNLP \cite{Lorenz_Pearson_Meichanetzidis_Kartsaklis_Coecke_2021}, and the multi-class classification results are a significant step forward. The calculated expressibility and entangling capability metrics indicated that the model has favorable properties \cite{Sim_Johnson_Aspuru_Guzik_2019}.

We have not yet addressed how to adapt the model to database changes. Since the training process is iterative, it could be utilized in retraining after the database has been updated. We will study how much the database needs to change before the model starts performing differently. Additionally, the current model was not trained or evaluated on real hardware. Due to its moderate size, deploying it on real hardware would be feasible. Studying the model's performance on real hardware will be part of the future research.

It is unlikely that quantum computing will take an instant ''quantum leap'' to outperform classical computing. Instead, it will undergo incremental improvements in both software and hardware, resulting in advantages in specific tasks. Our current work combines theoretical and experimental advances in applying quantum computing to databases, progressing us toward these goals.